\documentclass[12pt]{article}
\usepackage[round]{natbib}
\usepackage{amsmath,amssymb,amsfonts}
\usepackage{array}
\usepackage{caption}
\usepackage{setspace}
\usepackage{authblk}
\usepackage{graphicx}
\usepackage{booktabs}
\usepackage{siunitx}
\usepackage{hyperref}
\usepackage{float}
\usepackage{url}
\usepackage[flushleft]{threeparttable} 

\hypersetup{hidelinks}

\begin{document}

\title{Measuring economic outlook in the news\thanks{We would like to thank Philippe Goulet Coulombe, participants at the ECONDAT Fall 2025 Meeting, at the 21st Annual NBP-SNB Joint Seminar and at the KOF-SNB Workshop for their insightful and most valuable comments. We would also like to thank Milan Seifert and Andrin Gehrig for their research assistance. The views, opinions, findings, and conclusions or recommendations expressed in this paper are strictly those of the author(s). They do not necessarily reflect the views of the Swiss National Bank (SNB). The SNB takes no responsibility for any errors or omissions in, or for the correctness of, the information contained in this paper.\newline 
For this publication, use was made of media data made available via Swissdox@LiRI by the Linguistic Research Infrastructure of the University of Zurich (see \url{https://www.liri.uzh.ch/en/services/swissdox.html} for more information).}}

\author[1]{Elliot Beck}
\author[1]{Franziska Eckert}
\author[2,3]{Linus Kühne\thanks{Parts of this research were conducted while the author was employed at the Swiss National Bank.}}
\author[1]{Helge Liebert}
\author[1]{Rina Rosenblatt-Wisch}

\affil[1]{\small Swiss National Bank, Zurich}
\affil[2]{\small Seminar for Statistics, ETH Zurich}
\affil[3]{\small ETH AI Center, ETH Zurich}

\date{February 2026} 

\maketitle

\begin{abstract}
\noindent
We develop a resource-efficient methodology for measuring economic outlook in news
text that combines document embeddings with synthetic training data generated by
large language models. Applied to 27 million news articles, the resulting indicator
significantly improves GDP growth forecast accuracy and captures sentiment shifts
weeks before official releases, proving particularly valuable during crises. The indicator
outperforms both survey-based benchmarks and traditional dictionary methods
and is interpretable, allowing identification of specific drivers of economic sentiment.
Our approach addresses key institutional constraints: it performs sentiment classification
locally, enabling analyses of proprietary news content without transmission
to external services while requiring minimal computational resources compared to
direct large language model classification.
\end{abstract}


\vfill

\newpage

\section{Introduction}
\begin{spacing}{1.25}

Timely information is essential for informed decision-making during periods of crisis and elevated uncertainty. Traditional macroeconomic indicators--such as GDP growth, unemployment, and inflation--are typically released monthly or quarterly, limiting their usefulness in fast-moving economic environments. This release lag has motivated growing interest in alternative measurement approaches that exploit high-frequency data to provide real-time insights into economic conditions.

News articles represent a particularly valuable source of alternative data. The underlying intuition is that media coverage reflects real-time shifts in economic outlook and sentiment, often capturing developments well before they appear in official statistics or survey-based indicators \citep[e.g.,][]{tetlock:2007, bybee:2024}. A sudden policy announcement or market shock, for instance, generates immediate news coverage but may take weeks to surface in consumer confidence or business sentiment surveys. Recent advances in natural language processing have made it feasible to extract timely economic signals from news streams at scale, with studies demonstrating that such signals can meaningfully improve the precision of macroeconomic forecasts compared to traditional indicators alone \citep[e.g.,][]{hirschenbuehl:2021, woloszko:2024, kalamara:2022, ashwin:2024}.

Large language models (LLMs) have emerged as a powerful tool for extracting information from text \citep[e.g.,][]{ash:2025,kwon:2024,kwon:2025}. Unlike traditional keyword-matching approaches, LLMs capture complex semantic structures and contextual nuances in language. However, their application faces significant practical constraints. State-of-the-art models require substantial computational resources, whether through internal hosting infrastructure or external services--a challenge for many institutions. Moreover, data usage agreements (DUAs) from news providers often restrict transmitting data to external services, while internal infrastructure typically cannot support the most advanced models.

This paper develops an efficient methodology for measuring economic sentiment in news text that retains the advantages of state-of-the-art LLMs while addressing data confidentiality and resource constraints. We apply this methodology to construct the News-based Economic Outlook for Switzerland (NEOS), a novel indicator that extracts real-time economic sentiment from published news articles. NEOS offers three key advantages: a) it is highly resource-efficient--making it compatible with restrictive DUAs; b) it is interpretable--a crucial feature for policymaking; and c) it significantly improves GDP growth forecast accuracy.

Our approach uses a multi-step process that strategically combines different models to balance performance and efficiency. First, news articles are converted into high-dimensional document embeddings using a specialized, computationally efficient transformer model. Second, we use a state-of-the-art LLM to generate synthetic articles that mimic typical news article content while conveying a clearly positive or negative economic outlook. These synthetic articles are embedded and used to train a logistic regression model, which is then applied to real news article embeddings to produce economic outlook scores. The final NEOS indicator aggregates these scores across all articles in a given month. This architecture is highly resource-efficient: we estimate that the operating costs of our approach are several orders of magnitude lower compared to directly assessing economic outlook based on the same corpus of news articles using a commercial LLM, while maintaining the semantic understanding capabilities of advanced language models.

A critical practical advantage of our methodology is its compatibility with restrictive DUAs. Many news providers and media organizations impose strict limitations on how their content can be processed and stored, often prohibiting transmission to external servers or analysis through third-party commercial APIs. Our approach circumvents these constraints by performing all sentiment classification locally using embeddings and logistic regression, only generating synthetic training data with external LLMs. This design enables institutions to analyze proprietary or DUA-restricted news corpora without violating contractual obligations, substantially expanding the range of feasible applications. The methodology is particularly valuable for central banks, regulatory agencies, and research institutions that routinely work with sensitive or commercially restricted data sources.

Our methodology also addresses several limitations of the existing research. Previous studies have often analyzed sentiment from single outlets--such as the Wall Street Journal--or only a few outlets, potentially missing important dimensions of public sentiment \citep[e.g.,][]{tetlock:2007, garcia:2013}. We demonstrate meaningful added value from incorporating diverse publication sources: expert publications and broader media outlets capture different facets of sentiment, and combining both provides a more balanced assessment of economic outlook.

Furthermore, NEOS offers interpretability advantages over black-box classification approaches. By measuring each article's influence on the indicator and retrieving original content, we can identify specific drivers of economic outlook at any point in time and conduct sector-specific analyses that may reveal shifts in trade, financial markets, or other areas before they appear in official data.

We evaluate NEOS through a pseudo-out-of-sample forecasting study of GDP growth. Our results show that NEOS significantly outperforms both survey-based benchmark indicators and traditional dictionary-based approaches. Timelier versions of NEOS, computed from partial monthly data, capture sentiment shifts weeks before official releases, proving particularly valuable during times of crises. Unlike dictionary methods that rely on simple keyword matching, NEOS captures the semantic meaning of articles, enabling deeper understanding of economic outlook. These findings align with recent research emphasizing the limitations of dictionary-based sentiment classification: while computationally efficient, such methods lack accuracy compared to transformer models and large language models \citep[e.g.,][]{audrino:2024,kirtac:2024,klahn:2025}.\footnote{Previous studies by \citet{binsbergen:2024} and \citet{barbaglia:2025} demonstrated the value of integrating news-based indicators with machine learning techniques to improve forecast accuracy.}

Our work relates closely to recent studies leveraging LLMs for economic sentiment analysis and forecasting, though with important distinctions. \citet{kwon:2025} extracted growth and inflation sentiment indices by directly parsing press narratives using GPT-4 and GPT-5, decomposing aggregate sentiment into demand, supply, and structural drivers. While their approach improves forecasting performance, our method offers greater efficiency through embeddings that can run on on-premises systems without requiring large cloud-based infrastructure. \citet{buckmann:2025} similarly used logistic regression on embeddings from small generative models for sentiment analysis, highlighting advantages in privacy, availability, and cost. Our study extends this work through a two-sample approach using synthetic documents alongside real articles, benefiting from state-of-the-art LLMs while avoiding manual labeling. Finally, \citet{seiler:2025} computed an economic sentiment indicator from Swiss business tendency surveys using a BERT-based model. While that indicator tracks the business cycle well, our method leverages a richer, more diverse article corpus and incorporates synthetic documents to enable comparability.

The remainder of this paper proceeds as follows. Section \ref{sec:data} introduces our data. Section \ref{sec:methodology}  outlines the methodology for computing NEOS and extracting insights into its drivers. Section \ref{sec:results} evaluates NEOS's forecasting performance and illustrates key drivers of the indicator. Section \ref{sec:sensitivity} presents sensitivity analyses validating our methodological design choices. Section \ref{sec:conclusion} concludes the study.

\section{Data}
\label{sec:data}

We use the Swissdox@LiRI database, a collaboration between the Linguistic Research Infrastructure at the University of Zurich (LiRI) and Schweizer Mediendatenbank AG, to obtain a comprehensive and representative sample of Swiss news articles. The full database contains more than 29 million articles from over 260 Swiss media outlets, dating back to 1911. 

The Swissdox database covers almost all publications by Swiss media outlets in recent years, with the exception of Italian-language Swiss media. Therefore, our analysis focuses on articles published in German and French. We also discard some very specific media outlets that presumably do not contain any relevant articles (e.g., TV guides and various magazines). In addition, we only concentrate on print and online news articles, excluding radio and TV transcripts. Our analysis spans from January 1999 until November 2025. Thus, we use all news articles, print and online, in German and French, from 159 media outlets, amounting to roughly 27 million articles. The database is growing at a rate of approximately 100,000 articles per month, with new content typically becoming available two business days after publication.

We restrict our analysis to articles published from January 1999 onward to avoid the effects of compositional changes in the early years of the archive. The raw text contains markup to identify elements of an article (e.g., title, body, paragraphs, tables, boxes). During preprocessing, we remove all tables and boxes, remove all markup tags and concatenate the title and body of the article.

\begin{figure}
    \centering
    \includegraphics[width=\textwidth]{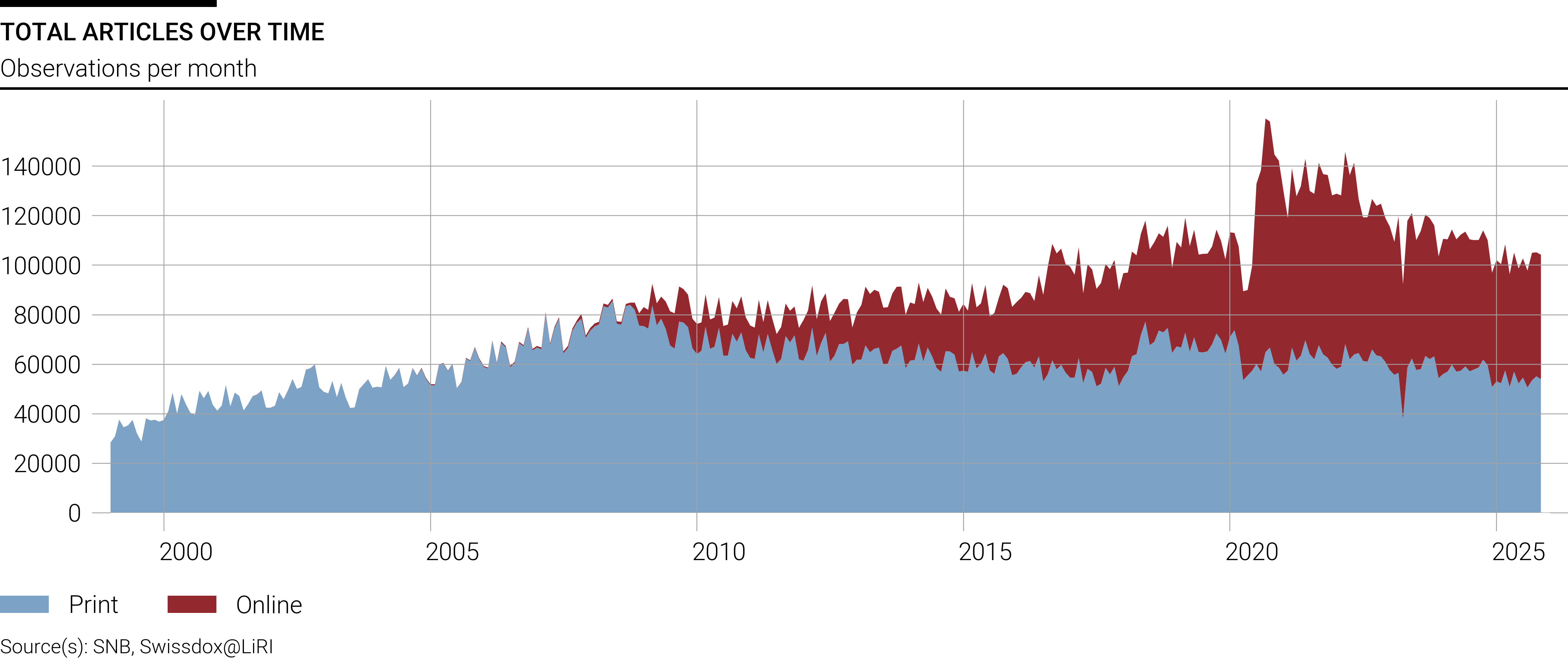}
    \caption{Number of all articles over time, broken down by publication type (print vs. online).}
    \label{fig:swissdox_all}
\end{figure}

Figure \ref{fig:swissdox_all} shows the total number of articles by publication type over time. The total number of articles has steadily increased, with a sharp increase in 2020, primarily due to the surge in online articles during the COVID-19 pandemic. This increase was driven by daily updates on COVID-19 case numbers, with new entries being added to the database regularly.

\section{Methodology}
\label{sec:methodology}
\begin{figure}
\centering
\includegraphics[width=\textwidth]{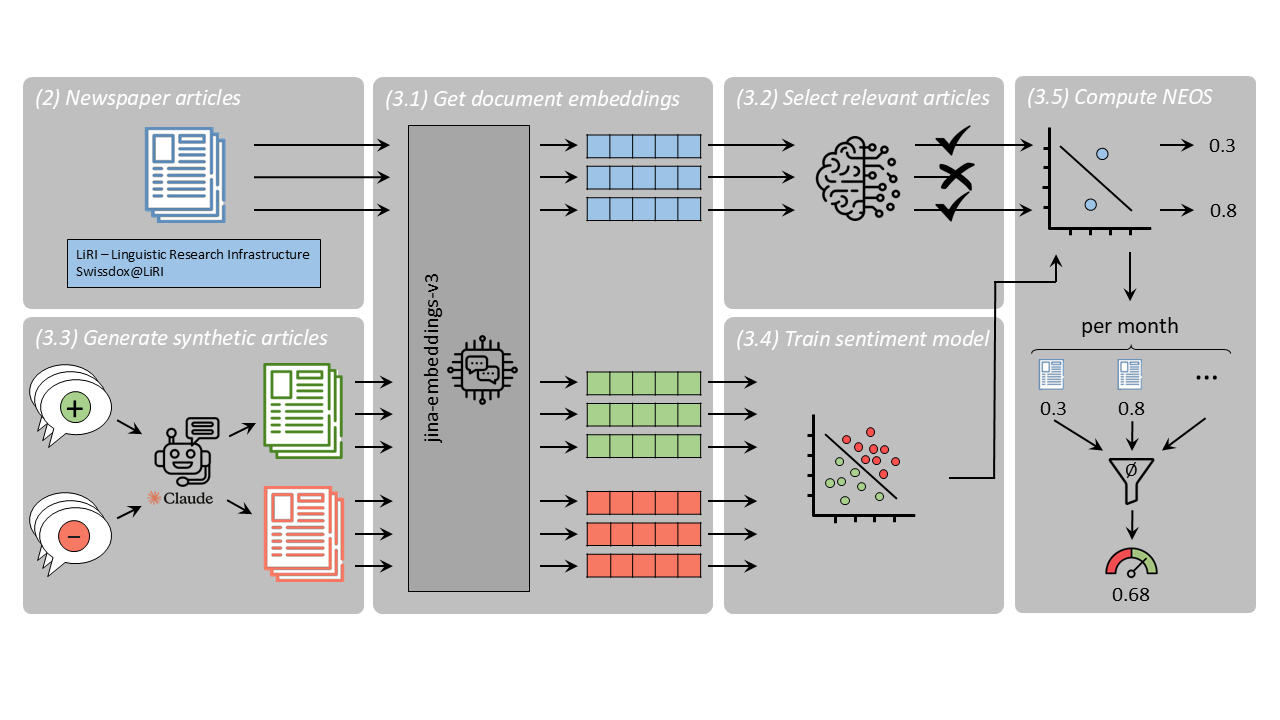}
\captionsetup{skip=-0.7cm}
\caption{Overview of the computation steps for NEOS.}
\label{fig:neos_overview}
\end{figure}

This section describes the individual steps involved in computing NEOS. Figure \ref{fig:neos_overview} gives an overview. We use an efficient and scalable two-sample estimation approach to generate the indicator. As illustrated in Figure \ref{fig:neos_overview}, we leverage both the Swissdox news articles (Section~\ref{sec:data}, i.e., the top stream in Figure~\ref{fig:neos_overview}), and synthetic articles generated by an LLM (Section~\ref{subsec:example_articles}, i.e., the bottom stream in Figure~\ref{fig:neos_overview}). We run both sets of articles through a small encoder-only embedding model to obtain the numeric representations (Section \ref{subsec:doc_embeddings}). Using these embeddings, we select articles from the Swissdox data that relate to the state of the economy (see Section \ref{subsec:filter_relevant}). We use only the synthetic data sample to train the sentiment model (Section \ref{subsec:logreg}). We then take the fitted model to predict the sentiment scores of the relevant articles in the Swissdox data and aggregate these scores to compute NEOS (Section \ref{subsec:compute_neos}). In addition, we decompose NEOS with different methods to ensure interpretability (Section \ref{subsec:interpret}). We detail each step below.

\subsection{Generate document embeddings}
\label{subsec:doc_embeddings}

To transform article texts into a numerical format suitable for analysis, we use the jina-embeddings-v3 model developed by JinaAI \citep{sturua:2025}. This encoder-only embedding model converts each article into a 1024-dimensional vector that captures the semantic meaning of the content, preserving the overall message and tone of the article. We choose jina-embeddings-v3 because the model is multilingual, i.e., capable of processing both German and French articles, and can handle relatively long documents with up to 8192 input tokens. Previous generation embedding models like BERT only support short input sequences of 512 tokens, which are insufficient to generate high quality embeddings of longer content like news articles. At the same time, jina-embeddings-v3 is still a comparatively small model with only 570M parameters, allowing for fast inference on moderate hardware.

Generating embeddings has several advantages compared to directly using LLMs to classify the news articles. Although large decoder-only models (GPT, Llama, etc.) can be used to classify sentiment in texts directly, these models typically contain tens or hundreds of billions of parameters, making them computationally expensive and slow for inference, whereas dedicated embedding models like jina-embeddings-v3 are significantly more compact and efficient. The embeddings preserve the semantic meaning, which means that the overall message and tone of a document is captured in the embeddings and can be used to build a dedicated sentiment model. 

We estimate that computing an economic outlook indicator similar to NEOS using ChatGPT-4o to directly analyze the economic outlook of all news articles would have cost over \$30,000 at the time that NEOS was developed. In contrast, NEOS operates at a much lower cost. It can be run on consumer-grade hardware, and if deployment on an external platform such as AWS Cloud were allowed, the computation of NEOS could be completed for approximately \$100, based on the costs at the time of development.\footnote{For development, we used a single local NVIDIA L40S GPU with 48GB of memory. Note that the Swissdox@LiRI data are subject to restrictive DUAs and cannot be analyzed on external platforms.}

In addition, large generative models, such as ChatGPT-4o, are commercial offerings which require data to be transmitted to external API endpoints. The licensing terms offered by the providers are typically irreconcilable with common data use agreements (DUAs) or institutional data protection requirements. This effectively prohibits processing sensitive or proprietary data. The news articles we use are subject to a restrictive DUA, so on-premises deployment is essential to maintain data sovereignty and avoid DUA violations. Licensing terms offered by other news data providers (Bloomberg, Reuters, Dow Jones Factiva) are similarly restrictive. The smaller footprint of jina-embeddings-v3 makes it feasible to deploy it on modest hardware infrastructure, while large generative models would require substantial GPU resources that are infeasible for most on-premises environments.

\subsection{Select relevant articles}
\label{subsec:filter_relevant}

To isolate the articles relevant to the state of the economy, we employ a classification approach. Some media outlets within the Swissdox@LiRI database provide information about the section in which an article was published. We assume that articles in economics-related sections like ``business'', ``markets'', and ``economics'' are likely to contain information about the state of the economy. We consider these economics-related articles as relevant, and articles published in other sections (e.g. ``sports'') non-relevant. We use the embeddings of the articles with known sections to train a neural network model to classify whether any given article is related to economics (i.e., relevant) or not.

We evaluate different candidate models to classify the relevant (economics-related) articles before settling on the neural network. To train the models, we use a balanced sample of 300,000 articles, randomly selecting 150,000 articles each from both the relevant and non-relevant articles. We then evaluate the performance of the classifiers on a test set, an additional set of 150,000 articles, which were not used during training, also equally split between relevant and non-relevant articles. In our model evaluation, we achieved approximately 92\% accuracy using a linear model (L2-penalized logistic regression). A random forest also achieved 92\% accuracy. We settle on a simple feedforward neural network with four hidden layers in a funnel structure (512, 256, 128 and 64 neurons), ReLU activation, L2 regularization, dropout, batch normalization and sigmoid output layer; achieving 95\% accuracy.

Next, we evaluate the precision-recall tradeoff. Figure \ref{fig:acc_prec} in the Appendix \ref{sec:accuracy vs precision} displays the overall accuracy vs.\ precision for different candidate models, illustrating our choice of model and classification threshold. To improve the model's ability to correctly identify relevant articles (precision) and limit the number of false positives, we increase the classification threshold from 0.5 to 0.8, trading off some accuracy for gains in precision. The final model achieves 93\% accuracy and 95\% precision on the test data, demonstrating the strong performance of our classifier. We then use the trained classifier to label and select all relevant articles in the full data.

\begin{figure}
\centering
\includegraphics[width=\textwidth]{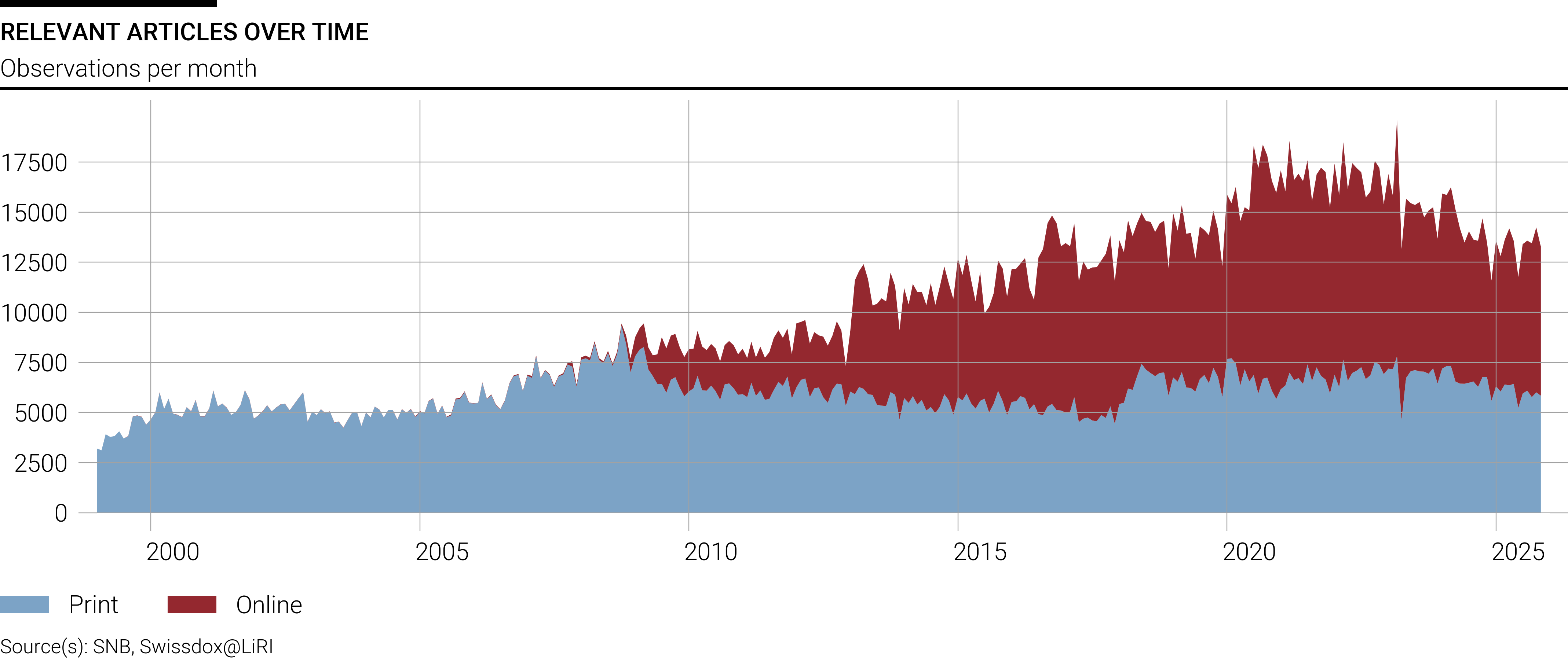}
\caption{Number of relevant articles over time, broken down by publication type (print vs. online).}
\label{fig:swissdox_relevant}
\end{figure}

Based on the model prediction, we select only relevant articles from the full sample, leaving us with a refined set of economics-related articles. In total, we retain 2.7 million relevant articles in German and 0.4 million relevant articles in French. The number of relevant articles per month increased over time, from approximately 3,000 articles in January 1999 to a peak of approximately 17,000 articles per month during the COVID-19 pandemic, and to approximately 13,000 articles in November 2025. Figure \ref{fig:swissdox_relevant} shows the distribution of these articles over time, segmented by publication type. By selecting only articles relevant to the state of the economy, we improve the signal in the data.

\subsection{Generate synthetic articles}
\label{subsec:example_articles}
To avoid the time-intensive, error-prone and subjective process of reading and manually labeling the sentiment of parts or even all relevant articles, we leverage a state-of-the-art LLM to generate synthetic articles conveying clearly positive or negative economic outlooks. The synthetic articles are created using Claude 3.5 Sonnet \citep{Anthropic2024}. We generate in total 256 synthetic articles, 128 articles of which convey a stereotypical positive economic outlook and 128 articles of which convey a stereotypical negative economic outlook. We instruct the LLM to generate the synthetic articles in such a way that they mimic a typical news article in style and structure. We specify that some synthetic articles should cover the economy in general, while others should cover specific parts of the economy, i.e., financial markets, the labor market, real estate market, international trade, consumption, business situation and macroeconomic outlook. An example prompt, along with a positive and negative synthetic article, is provided in Appendix \ref{sec:prompt_examples}. We assign a score of one to the synthetic articles with a positive outlook and a score of zero to the synthetic articles with a negative outlook. 

\begin{figure}
\centering
\includegraphics[width=\textwidth]{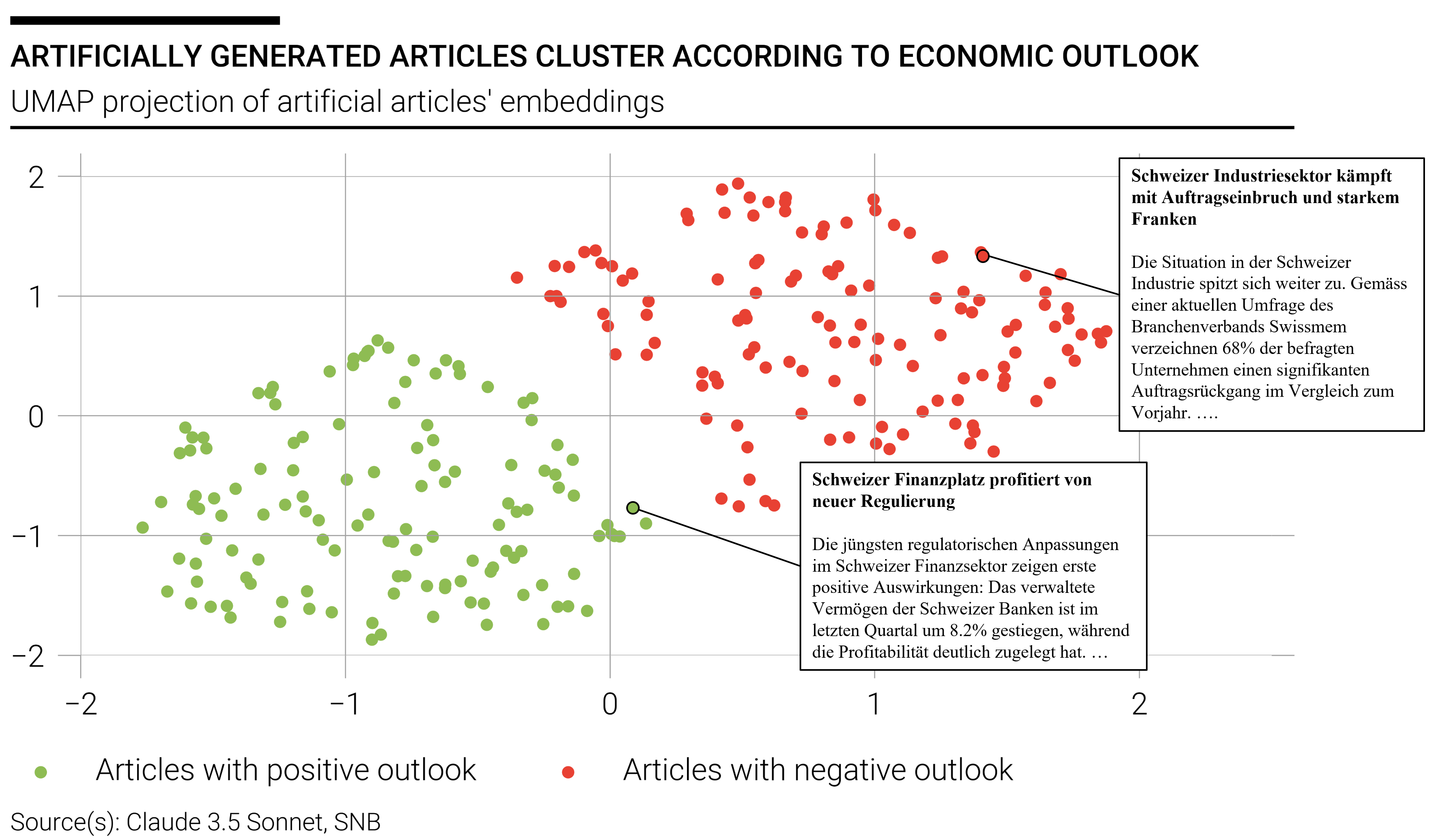}
\caption{UMAP visualization of the synthetic articles with positive and negative economic outlooks.}
\label{fig:umap}
\end{figure}

The synthetic articles are then converted into document embeddings using the jina-embeddings-v3 model described in Section \ref{subsec:doc_embeddings}. For exploratory analysis and visualization, we apply UMAP \citep{mcinnes:2020}, a dimensionality reduction approach, to reduce the high-dimensional embeddings to two dimensions. Figure \ref{fig:umap} plots the two-dimensional representation of the 256 synthetic articles. We color the articles with positive outlook green and the ones with negative outlook red. We observe two distinct clusters corresponding to positive and negative outlook. This separation confirms that the embeddings effectively capture sentiment.

Using synthetic articles has several advantages over manual labeling of real articles. Human labeling can be error-prone, difficult to reproduce, and time-intensive. When performed correctly, it is also costly, as it requires multiple individuals to independently label a large number of articles to validate the alignment of the labels. In contrast, generating synthetic articles is cheap and produces higher quality training data, since real world articles are rarely fully positive or negative, but typically contain ambiguous wording. The use of synthetic articles allows us to clearly assign one or zero as labels to the articles and no arbitrary values in between are required.

\subsection{Train the sentiment model}
\label{subsec:logreg}

The clear separation of the synthetic articles into two distinct clusters in Figure \ref{fig:umap} suggests that a linear classification model can be used to distinguish between positive and negative economic outlook. To achieve this, we fit a logistic regression model to the embedded synthetic articles. Given that the dimensionality of the embeddings (1024 features) exceeds the number of observations (256 articles), we apply regularized logistic regression \citep{cessie:1992} to prevent overfitting.

\subsection{Measure economic outlook}
\label{subsec:compute_neos}

Once the logistic regression model is trained, we apply it to the embeddings of the relevant articles identified in Section \ref{subsec:filter_relevant}. The model generates a probability score for each article, measuring the likelihood that the article reflects a positive economic outlook. A score closer to one indicates a positive outlook, while a score closer to zero indicates a negative outlook.

To compute the final indicator, we aggregate the probability scores of all relevant articles within each month. This provides a monthly measure of the overall sentiment conveyed by the media. In addition, we compute timelier versions of NEOS aggregating the probability scores of the relevant articles published within the first 7, 14, or 21 days of each month. This allows for the generation of sentiment signals up to three weeks ahead of traditional survey-based economic indicators.

\subsection{Interpretability of NEOS}
\label{subsec:interpret}

NEOS is transparent and interpretable, which is essential for policymaking. We can measure the influence of each article on our indicator and retrieve its original content. This enables us to gain insights into what drives the economic outlook at any given time. Inspecting individual articles can be useful to explore what drives NEOS at a specific point in time. However, to interpret NEOS systematically over time, this approach is infeasible. We therefore aggregate articles within broader topics. We then decompose NEOS by topics, calculating the contribution of each topic to the overall indicator as follows:  
\begin{equation*}
 x_{t} = \sum_{i=1}^{N_t} x_{t, i} =  \sum_{i=1}^{N_t} \sum_{j=1}^{T} p_{t, i, j} x_{t, i} = \sum_{j=1}^{T} \sum_{i=1}^{N_t}  p_{t, i, j} x_{t, i} = \sum_{j=1}^{T} c_{t, j},  
\end{equation*}
where $x_{t}$ is NEOS in period $t$, $x_{t,i}$ is the score of article $i$ in period $t$, $N_t$ is the number of articles in the respective period, $T$ is the number of topics, $p_{t,i,j}$ is the probability than article $i$ belongs to topic $j$ in period $t$ (with $\sum_{j=1}^{T} p_{t, i, j} =1$) and $c_{t, j}$ is the contribution of topic $j$ to $x_t$ in period $t$. Note that when decomposing the monthly standardized version of NEOS, the same mean and variance of the monthly indicator should be used to center and scale $x_{t,i}$ per article.

The key element in this approach is how to determine the topics and the corresponding probabilities for each article. There are several possible approaches, each suitable for different use cases. In the following, we discuss three selected approaches.

\paragraph{Keyword-based} A keyword-based method is useful to investigate the influence of specific ad-hoc topics, e.g., to quickly validate a policymaker's intuition. By defining one or more keywords that must be present in an article for it to be considered relevant to a particular topic, we obtain $p_{t, i, j} \in \{0,1\}$. In this simple form, this only allows us to distinguish between articles that belong to the defined topic and those that do not. Matching keywords is a crude approach to identify topics--articles which contain only one relevant section or where the entire article is relevant are weighed equally. Additionally, keyword-based methods lack context awareness which can lead to incorrect interpretations.

\paragraph{Classification-based}Classification-based methods are useful for predefined topics. A common approach is to label a subset of the data by humans or LLMs for training \citep[e.g.][]{bucur:2025}. To define the topics, expert knowledge is required (see Section~\ref{subsec:example_articles} for a discussion of the advantages and disadvantages of manual labeling). The topic labels are then used to train a classifier, using the embeddings as input. We exploit the fact that we instructed the LLM to generate synthetic articles for different economic sectors (labor market, trade and financial markets; see Section~\ref{subsec:example_articles}). Thus, our synthetic articles are already categorized by economic sectors. These categories can serve as topics. We train a regularized multinomial logistic regression on the embeddings of the synthetic example articles and apply the trained model to the embeddings of the news articles to obtain $p_{t, i, j}$. To simplify and ease interpretability, we limit the number of topics assigned to each article: We limit the total number of topics to a maximum of three and select only the topics with the highest probabilities for which the cumulative sum of their probabilities in decreasing order exceeds 0.7. We rescale the probabilities of the selected topics per article such that they sum up to one again.

\paragraph{Clustering-based}Both of the above approaches rely on pre-defined topics. In contrast, automatically generated topics allow monitoring emerging trends in the data. To achieve this, we first apply UMAP \citep{mcinnes:2020} to the embeddings of the news articles to obtain a new set of embeddings with lower dimensionality. Thereafter, we fit K-means clustering to the lower-dimensionality embeddings of the latest available month, i.e., November 2025. We then apply the resulting model to the historical data. This allows us to capture the latest developments while ensuring a consistent history. We select K-means over more sophisticated clustering algorithms as the algorithm is deterministic (for a fixed seed), scalable and able to predict the cluster membership for new data points consistently.

\paragraph{}To gain a deeper understanding of the drivers of a topic during a specific time, it is possible to retrieve individual articles (e.g., most positive/negative, highest probability for specific topic, highest positive/negative contribution within topic), generate word clouds or use an LLM to generate summaries of the articles within the topic and time range. These methods for enhanced interpretability can be applied to topics identified by all three approaches but are particularly important for automatically generated topics, due to the absence of prior knowledge about these topics and their content.

We show how all of these different approaches can be used in practice in Section \ref{sec:contributions}.

\section{Results}
\label{sec:results}
\begin{figure}[t]
    \centering
    \includegraphics[width=\textwidth]{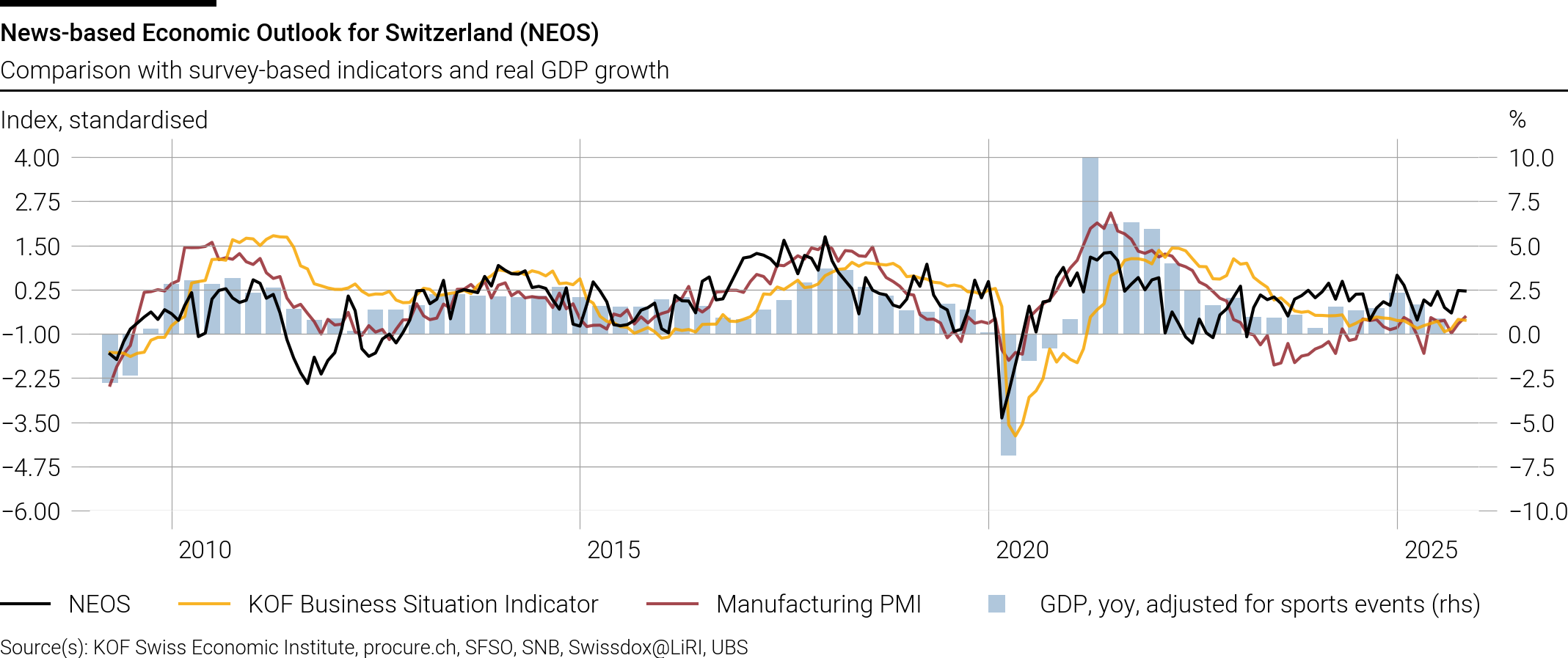}
    \caption{NEOS and benchmark indicators together with Swiss real GDP growth (year-on-year, adjusted for sports events; right-hand scale). All indices are standardized.}
    \label{fig:neos}
\end{figure}

Figure~\ref{fig:neos} shows monthly NEOS alongside real quarterly year-on-year GDP growth and two traditional, monthly, survey-based leading indicators (KOF Business Situation Indicator and Manufacturing Purchasing Managers' Index). NEOS closely tracks GDP growth and captures turning points similarly to the established indicators. To demonstrate NEOS's leading properties, we analyze the correlations between NEOS and GDP growth at different lags. Appendix ~\ref{sec:correlations} shows the correlations alongside the same correlations for other indicators, which are based on surveys or news text, for comparison. The correlations for NEOS are consistently high, often surpassing those of the other indicators. Next, we assess whether this strong co-movement of NEOS with GDP growth translates into improved forecasting performance.

In the remainder of this section, we conduct a comprehensive forecast evaluation of NEOS. We first compare its out-of-sample forecasting performance against a wide range of alternative indicators. We then examine the importance of NEOS's timeliness by assessing how its predictive accuracy varies with different publication lags. Next, we analyze NEOS's performance over time, highlighting how timely information helps during periods of crisis and high volatility. We also investigate our indicator's incremental predictive power by testing whether NEOS adds significant information beyond established indicators in multivariate forecasting models. 

In addition, we demonstrate how NEOS can be interpreted through an analysis of the underlying news content and sentiment drivers.

\subsection{Forecast evaluation}
\label{subsec:forecast_evaluation}
Following the approach of \citet{kalamara:2022}, we evaluate the ability of NEOS and other survey- or text-based indicators to improve real GDP growth forecasts in a pseudo-out-of-sample evaluation over the period from Q1~1999 to Q1~2025.\footnote{Since recent releases of GDP are often subject to backward revisions, we limit the analysis to Q1~2025.} The set-up is designed to mimic the information set of a policymaker: Towards the end of quarter~$t$, the policymaker wants to forecast year-on-year GDP growth for the current quarter $t$ and for the next two quarters $t+1$ and $t+2$, while official GDP data are available only up to $t-1$.

For each horizon $h \in \{0,1,2\}$, we estimate the following regression model
\begin{equation}
\label{eq:regression_model}
y_{t+h} \;=\; \alpha \,+\, \beta y_{t-1} \,+\, \gamma x_{t}^{(m)} \,+\, \varepsilon_t,
\end{equation}
where $y_t$ denotes the year-on-year GDP growth in quarter~$t$, $x_t^{(m)}$ is the average of the indicator values from the first~$m$~months of quarter~$t$, and $\varepsilon_t$ is a zero-mean error term. We compare our indicator against a range of established monthly indicators. These include widely used survey-based measures: the manufacturing and services Purchasing Managers’ Indices (PMIs) for Switzerland, the KOF Business Situation Indicator, and the SECO Consumer Sentiment Index.

In addition, we evaluate NEOS against two alternative text-based indicators derived from the same underlying news data. Both alternative indicators use a lexicon/dictionary-based approach. The first is the well-established Economic Policy Uncertainty (EPU) index proposed by \citet{baker:bloom:davies:2016}. The second follows the methodology of \citet{barbaglia:2025} and relies on a German translation of their sophisticated lexicon. Both of these text-based indicators serve as a natural benchmark: they rely on exactly the same underlying news data but employ a more straightforward counting approach (see Appendix~\ref{sec:lexicon} for details).

All survey indicators are released at the end of each month or shortly thereafter. To mimic the information set available to a policymaker near the end of quarter~$t$, we define $x_t^{(m)}$ as the average of the indicator values from the first two months ($m=2$) of quarter~$t$. Thus, forecasts of GDP growth $y_{t+h}$ rely on GDP observations up to quarter $t-1$ and indicator data available through the second month of quarter~$t$.

A key advantage of NEOS is that it can be continuously updated as new articles become available. To evaluate the benefits of this feature, we construct variants that include articles published in the first 7, 14, or 21 days of the third month of quarter~$t$ (see Section~\ref{subsec:compute_neos} for details). These variants incorporate more recent information and can be computed well before the end of the month—substantially earlier than conventional indicators. For these timelier variants of NEOS, we therefore set $m=3$, i.e. we use the first two months and additional indicator information available through the third month of quarter~$t$ in the forecasting exercise. This allows us to quantify the improvement in the forecast accuracy arising from NEOS's ability to exploit intra-month news flow and provide updates long before most traditional indicators are released.

To assess the forecast performance for each indicator, we compare the regression model in Equation~\eqref{eq:regression_model} to an autoregressive benchmark that only includes lagged GDP, i.e., Equation \eqref{eq:regression_model} with $\gamma = 0$. This corresponds to an AR(1) model for year-on-year GDP growth. For each indicator and forecast horizon, we compute the mean absolute error (MAE) of the pseudo-out-of-sample forecasts and report the ratio
\[
\frac{\text{MAE of Eq.\ \eqref{eq:regression_model} with indicator}}{\text{MAE of AR(1)}}.
\]
A value below one indicates that including the indicator as a predictor improves the accuracy of the forecast relative to the AR(1) model. We re-estimate the models using an expanding window, starting with an initial window containing the first eight quarters of data. Each forecast is based only on information that would have been available at the time it was made.

We assess the statistical significance of the improvements in forecast accuracy using a modified Diebold–Mariano test.\footnote{We follow \citet{beck:2026} and implement a modified version of the \citet{diebold:1995} test, using heteroskedasticity- and autocorrelation-consistent (HAC) standard errors to account for serial correlation in the forecast errors.} Table~\ref{tab:gdp_forecast} reports the MAE ratios, with the Diebold–Mariano p-values in parentheses. The lowest MAE ratio for each horizon is printed in bold.

\begin{table}[t]
\begin{center}
\caption{Forecasting Swiss year-on-year GDP growth from Q1~1999 to Q1~2025}
\label{tab:gdp_forecast}
\begin{tabular}{lccc}
\toprule
\textbf{Indicator} & \textbf{$h=0$} & \textbf{$h=1$} & \textbf{$h=2$} \\
\midrule
NEOS & 0.88 (0.03) & 0.78 (0.06) & 0.80 (0.05) \\
NEOS (first 7 days in the third month) & \textbf{0.87} (0.03) & 0.79 (0.06) & 0.80 (0.05) \\
NEOS (first 14 days in the third month) & 0.87 (0.03) & 0.78 (0.06) & 0.79 (0.04) \\
NEOS (first 21 days in the third month) & 0.88 (0.03) & \textbf{0.78} (0.06) & \textbf{0.77} (0.03) \\
\midrule
EPU for Switzerland & 1.12 (0.09) & 1.11 (0.89) & 1.01 (0.52) \\
Lexicon-based approach & 1.00 (0.49) & 0.97 (0.41) & 0.93 (0.30) \\
KOF Business Situation Indicator & 1.06 (0.86) & 1.01 (0.67) & 1.00 (0.48) \\
SECO Consumer Sentiment Index & 1.09 (0.80) & 1.01 (0.53) & 0.99 (0.47) \\
Manufacturing PMI for Switzerland & 0.94 (0.22) & 0.85 (0.16) & 0.84 (0.09) \\
Services PMI for Switzerland & 0.89 (0.10) & 0.95 (0.10) & 0.92 (0.12) \\
\bottomrule
\end{tabular}
\end{center}
\footnotesize
Notes: MAE ratios with the Diebold–Mariano p-values in parentheses. The best MAE values in bold. Due to limited data availability, the evaluation for the KOF Business Situation Indicator is limited to Q1~2009--Q1~2025 and that for the Services PMI is limited to Q1~2014--Q1~2025. In Appendix \ref{sec:forecast_eval_subperiods}, Tables \ref{tab:MAE_2009} and \ref{tab:MAE_2014}, we provide the results for the subperiods Q1~2009--Q1~2025 and Q1~2014--Q1~2025. Since the SECO Consumer Sentiment Index is a quarterly indicator, we regress GDP in $t+h$ on the indicator for quarter~$t$. Data sources: KOF Swiss Economic Institute, procure.ch, SECO, SNB, Swissdox@LiRI, UBS.

\end{table}

NEOS and its intra-month updated versions substantially improve forecast accuracy across all horizons, with MAE ratios ranging from 0.77 to 0.88 (implying error reductions of 12--23\% relative to the AR(1) benchmark). These gains are statistically significant at the 10\% level or better, as indicated by p-values below 0.06 in all cases.

The baseline NEOS achieves MAE ratios of 0.88 ($h=0$), 0.78 ($h=1$), and 0.80 ($h=2$). Incorporating news from the early days of the third month yields further improvements at longer horizons: the version using the first 14 days records 0.78 at $h=1$ and 0.79 at $h=2$, whereas the version using the first 21 days delivers the best performance at $h=1$ (0.78) and $h=2$ (0.77). At the nowcast horizon ($h=0$), the 7-day version is marginally superior (0.87). Overall, the intra-month variants match or outperform the baseline NEOS, demonstrating that substantial forecasting gains can be realized several weeks ahead of conventional indicator releases.

Relative to alternative indicators, NEOS dominates. The Economic Policy Uncertainty (EPU) index consistently yields MAE ratios above one. The alternative lexicon-based approach also yields MAE ratios close to one (0.93–1.00), with no statistically significant improvements. This highlights the value of NEOS's embedding-based methodology and synthetic labeling procedure, which extracts substantially richer macroeconomic signals from the same underlying news articles.

The KOF Business Situation Indicator and SECO Consumer Sentiment Index also generally perform worse than the AR(1) benchmark (MAE ratios around or above one) and show no significant forecasting power. The Manufacturing and Services PMIs deliver moderate improvements (MAE ratios of 0.84-0.95) but perform worse than NEOS across all horizons. Notably, PMI surveys are resource-intensive and follow fixed release schedules, whereas NEOS leverages readily-available news data and can be updated continuously.

Due to limited data availability, the evaluation for the KOF Business Situation Indicator is limited to Q1~2009--Q1~2025 and for the Services PMI to Q1~2014--Q1~2025. In Appendix \ref{sec:forecast_eval_subperiods}, Tables \ref{tab:MAE_2009} and \ref{tab:MAE_2014}, we provide the results for the respective subperiods. Overall, our findings carry over to these subsamples. 

Taken together, the results establish that our methodology is highly competitive, combining strong predictive performance with exceptional timeliness. 

\subsection{Importance of timeliness}
\label{sec:importance_of_timeliness}
\begin{figure}[t]
    \centering
    \includegraphics[width=\textwidth]{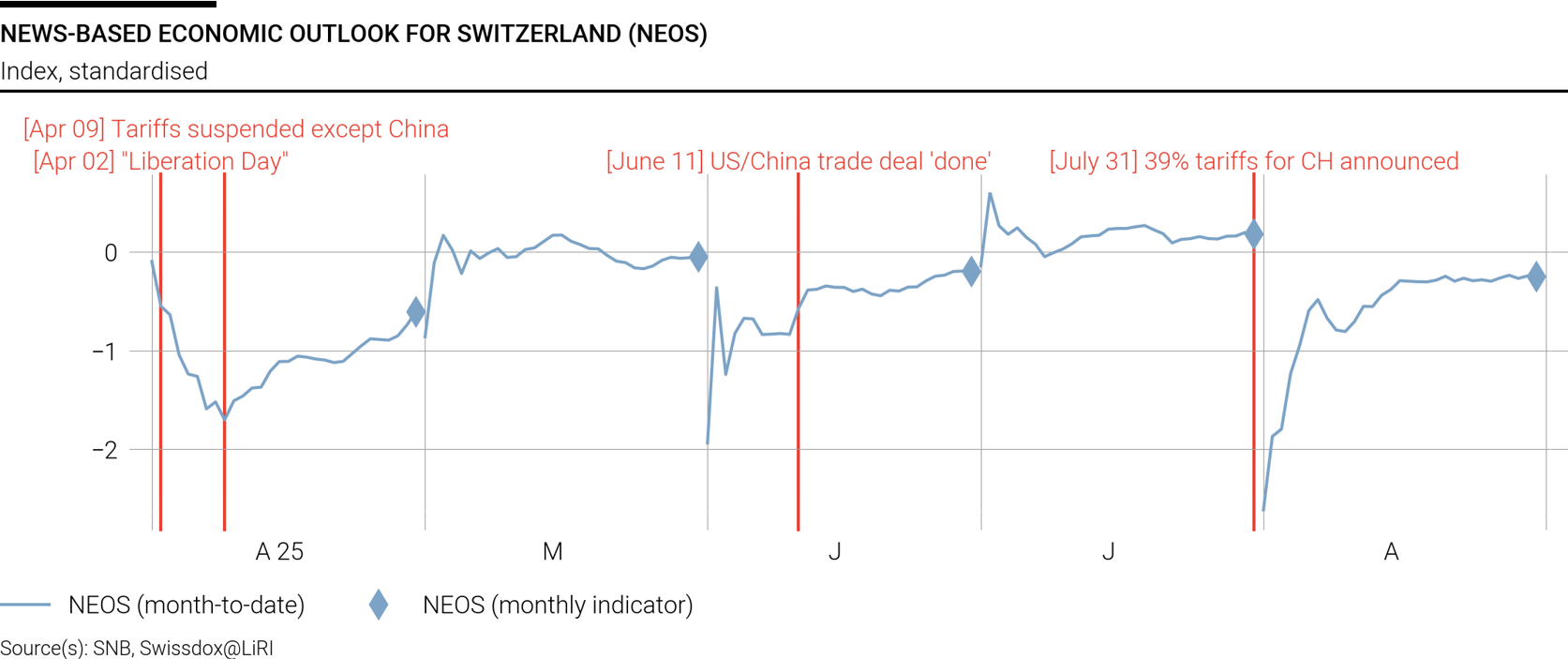}
    \caption{Daily month-to-date evolution of NEOS between April and August 2025.}
    \label{fig:neos_real_time}
\end{figure}

To illustrate the practical importance of timeliness, especially during times of crises, Figure~\ref{fig:neos_real_time} shows the daily month-to-date evolution of NEOS between April and August~2025. The daily month-to-date evolution is calculated by averaging all scores available from the first day of the month up to the current date. The blue diamonds mark the month-end values, which correspond to the monthly series shown in Figure~\ref{fig:neos}. The figure illustrates how sentiment in Switzerland shifted with US tariffs announcements. A policymaker monitoring NEOS at a daily frequency could have observed these shifts well before survey-based indicators or GDP releases became available.

\subsection{Performance over time}
\label{sec:performance_over_time}
Table \ref{tab:gdp_forecast} shows that including NEOS in the regression model in Equation \eqref{eq:regression_model} improves the accuracy of GDP growth forecasts relative to the AR(1) benchmark. To determine whether these gains are persistent or driven by outliers, we examine the path of the cumulative absolute error difference (CAED). 

For a given horizon $h$, the CAED at time $T$ is defined as:
\begin{equation*}
\mathrm{CAED}_{T,h}
=
\sum_{t=t_0}^{T}
\left(
\lvert e_{t,h}^{\mathrm{NEOS}} \rvert
-
\lvert e_{t,h}^{\mathrm{AR}} \rvert
\right),
\end{equation*}
where $|e_{t,h}^{\mathrm{NEOS}}|$ and $|e_{t,h}^{\mathrm{AR}}|$ are the absolute forecast errors of the NEOS-augmented regression and the AR(1) benchmark, respectively. The CAED is computed over the same pseudo out-of-sample evaluation period used in Section \ref{subsec:forecast_evaluation}. Under this specification, a downward-sloping trajectory indicates that the NEOS-augmented model outperforms the benchmark. In an ideal setting, the CAED series should be strictly downward sloping and remain below the zero threshold, indicating that the augmented model consistently incurs smaller forecast errors than the benchmark in every period.

\begin{figure}[ht!]
    \centering
     \includegraphics[width=\textwidth]{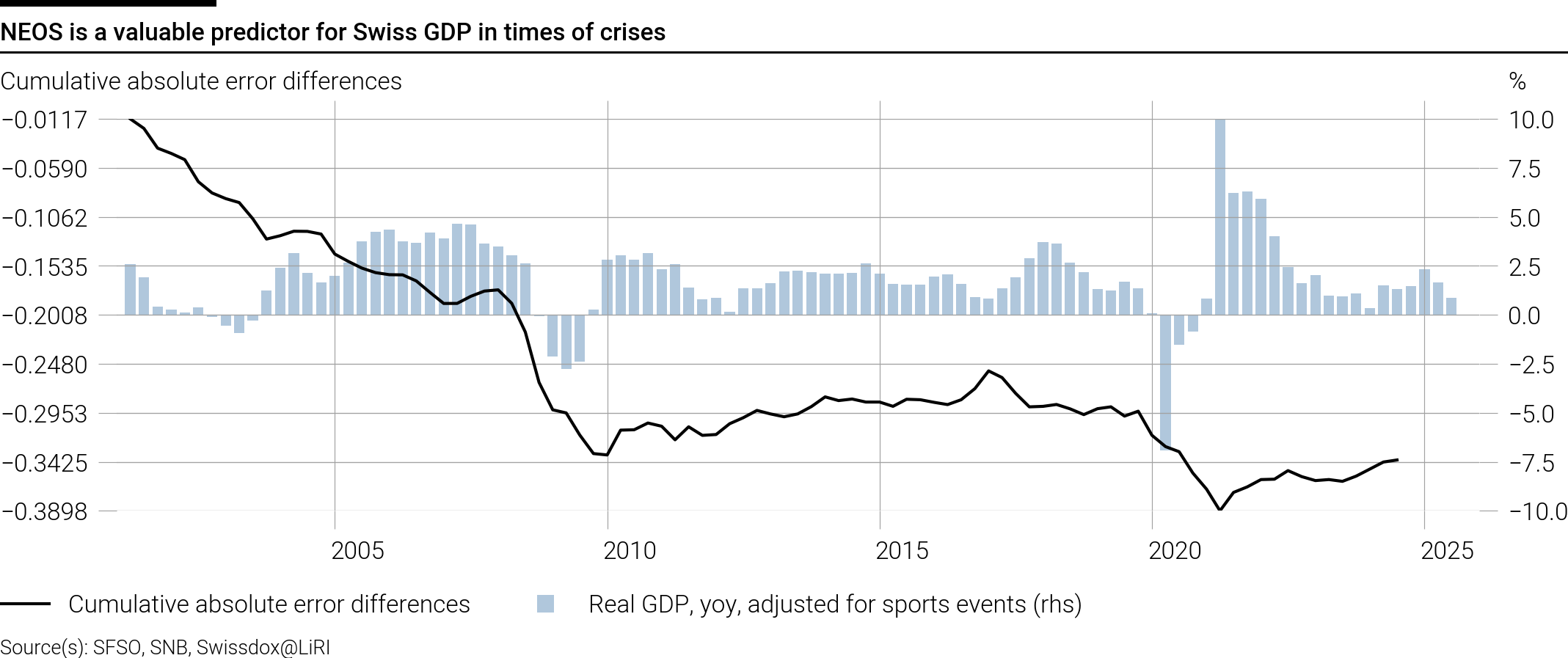}
    \caption{Cumulative forecast improvements compared to an AR(1) model for forecast horizon $  h=2  $.}
    \label{fig:crisis}
 \end{figure}

Figure \ref{fig:crisis} displays the CAED for horizon $  h=2  $. The series declines most sharply during periods of macroeconomic distress, such as the Great Financial Crisis and the COVID-19 pandemic. Outside of these crisis episodes, the series continues to trend downward overall but features several extended flat segments. These flat periods predominantly coincide with intervals of relatively stable and constant GDP growth rates, during which the simple AR(1) benchmark is particularly difficult to outperform. In such low-volatility environments, the NEOS-augmented model delivers predictive accuracy roughly on par with the AR(1), yielding no material additional gains. Nevertheless, the model still provides meaningful predictive improvements during more turbulent periods, and the overall pattern of outperformance remains visible across the full sample. While Figure \ref{fig:crisis} focuses on $  h=2  $ the results for horizons $  h=0  $ and $  h=1  $ are provided in Appendix \ref{sec:caed}; the qualitative patterns remain highly consistent across all three horizons.

\subsection{Kitchen sink regression}
To assess the incremental predictive power of the NEOS indicator, we perform a "kitchen sink" regression, including all established benchmarks in the model simultaneously. This test determines whether NEOS captures unique economic signals that are not already present in a linear combination of standard survey- and text-based measures. The regression equation is specified as follows:
\begin{equation*}y_{t+h} = \alpha + \gamma \text{NEOS}_{t} + \beta_1 y_{t-1} + \beta_2 \text{mPMI}_t + \beta_3 \text{SECO}_t + \beta_4 \text{LEX}_t + \beta_5 \text{EPU}_t + \epsilon_t.\end{equation*}

The control variables represent those indicators available over the full sample period and introduced in Section \ref{subsec:forecast_evaluation}: the manufacturing Purchasing Managers' Index (\textit{mPMI}), the SECO Consumer Sentiment Index (\textit{SECO}), and two text-based benchmarks derived from the same data source—the Economic Policy Uncertainty index (\textit{EPU}) and a German-translated lexicon-based measure (\textit{LEX}) following \citet{barbaglia:2025}.
The primary objective is to test the null hypothesis $$H_0: \gamma = 0.$$ If we reject this hypothesis, it implies that the NEOS indicator provides statistically significant predictive value beyond the information contained in traditional surveys and existing dictionary-based text methods.

The inclusion of the COVID-19 period presents a significant challenge for stable parameter estimation. The extreme fluctuations in GDP growth during 2020 and 2021 act as massive outliers that can disproportionately influence the regression coefficients, potentially masking the underlying structural relationship between the indicators and economic activity. Following \citet{schorfheide:song:2024}, who argued that excluding crisis observations is a robust and transparent alternative to complex outlier modeling in VAR contexts, we present the results both with and without the pandemic period to ensure the reliability of our estimates.

The results, presented in Table \ref{tab:kitchen_sink}, broadly correspond to the earlier univariate findings but offer deeper insight into the indicator's lead-lag structure. At the immediate horizon ($h=0$), the performance gap between NEOS and established indicators, specifically the manufacturing PMI, is relatively small, though the coefficient remains statistically significant when the COVID-19 period is excluded.

However, as the forecast horizon increases to $h=1$ and $h=2$, the relative strength of the NEOS indicator becomes more pronounced. Both the magnitude of the coefficient $\hat{\gamma}$ and its statistical significance increase at these longer horizons. This suggests that NEOS is particularly effective as a forward-looking measure; while survey indicators may reflect the current "nowcast" environment, NEOS appears to capture underlying economic sentiment that anticipates shifts in GDP growth several quarters ahead. These results remain robust even when accounting for the high-volatility COVID-19 period, confirming that NEOS is a valuable and distinct addition to the Swiss economic forecasting toolkit.

\begin{table}[t]
\centering
\begin{threeparttable}
\caption{Incremental predictive power of the NEOS: kitchen sink regression results}
\label{tab:kitchen_sink}
\newcolumntype{C}{>{\centering\arraybackslash}p{40pt}}
\begin{tabular}{l CCCCCC} 
\toprule
 & \multicolumn{2}{c}{$h=0$} & \multicolumn{2}{c}{$h=1$} & \multicolumn{2}{c}{$h=2$} \\
\cmidrule(lr){2-3} \cmidrule(lr){4-5} \cmidrule(lr){6-7}
$\hat{\gamma}$ & -0.001 & 0.002\textsuperscript{*} & 0.004\textsuperscript{*} & 0.005\textsuperscript{***} & 0.006\textsuperscript{***} & 0.005\textsuperscript{***} \\[-0.5ex]
                 & (0.002) & (0.001) & (0.002) & (0.001) & (0.002) & (0.001)  \\
\midrule
Other Indicators & \checkmark & \checkmark & \checkmark & \checkmark & \checkmark & \checkmark \\
\midrule
Include COVID-19 & \checkmark & & \checkmark & & \checkmark & \\
\bottomrule
\end{tabular}

\begin{tablenotes}
    \footnotesize
    \item Notes: This table reports the estimated coefficient $\hat{\gamma}$ for the NEOS indicator in the kitchen sink regression. The model controls for lagged GDP growth, mPMI, SECO consumer sentiment, and text-based benchmarks (EPU and LEX). HAC standard errors are reported in parentheses. The results are presented including/excluding the COVID-19 period (Q1 2020 until Q4 2021).  \textsuperscript{***}$p<0.01$, \textsuperscript{**}$p<0.05$, \textsuperscript{*}$p<0.1$.
\end{tablenotes}
\end{threeparttable}
\end{table}

\subsection{Interpreting NEOS}
\label{sec:contributions}
In the following, we analyze what drives NEOS. We therefore show the contributions of the topics determined by the three approaches presented in Section~\ref{subsec:interpret} as an example of their relevance in practice.

\begin{figure}[ht]
    \centering
    \includegraphics[width=\textwidth]{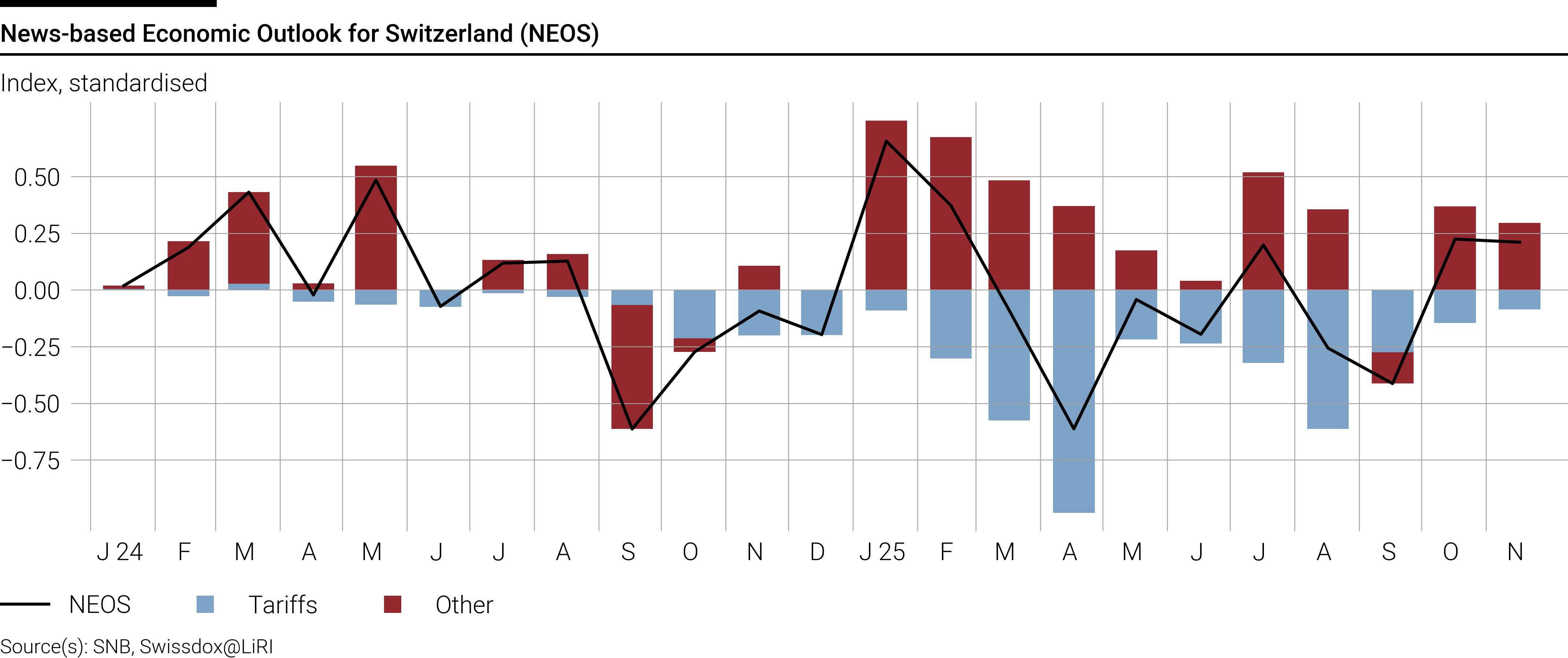}
    \caption{Contribution of the topic ``tariffs'' to NEOS as identified based on keywords.}
    \label{fig:neos_contr_keyword}
\end{figure}

Figure~\ref{fig:neos_contr_keyword} shows the contributions calculated using the keyword-based approach to determine $p_{t,i,j}$ for articles that contain the German or French word for ``tariff''. All the remaining articles that do not contain the keyword ``tariff'' are summarized in the category ``other''. In October 2024, the topic ``tariffs'' began to contribute markedly negatively to NEOS. At that time, former President Donald Trump had pledged during his election campaign to impose additional tariffs if he were elected. The negative contribution peaked on President Trump's ``Liberation Day'' in April 2025. It was also pronounced in August~2025 following the announcement of 39\% additional tariffs on Swiss goods on July 31.

\begin{figure}[ht]
    \centering
    \includegraphics[width=\textwidth]{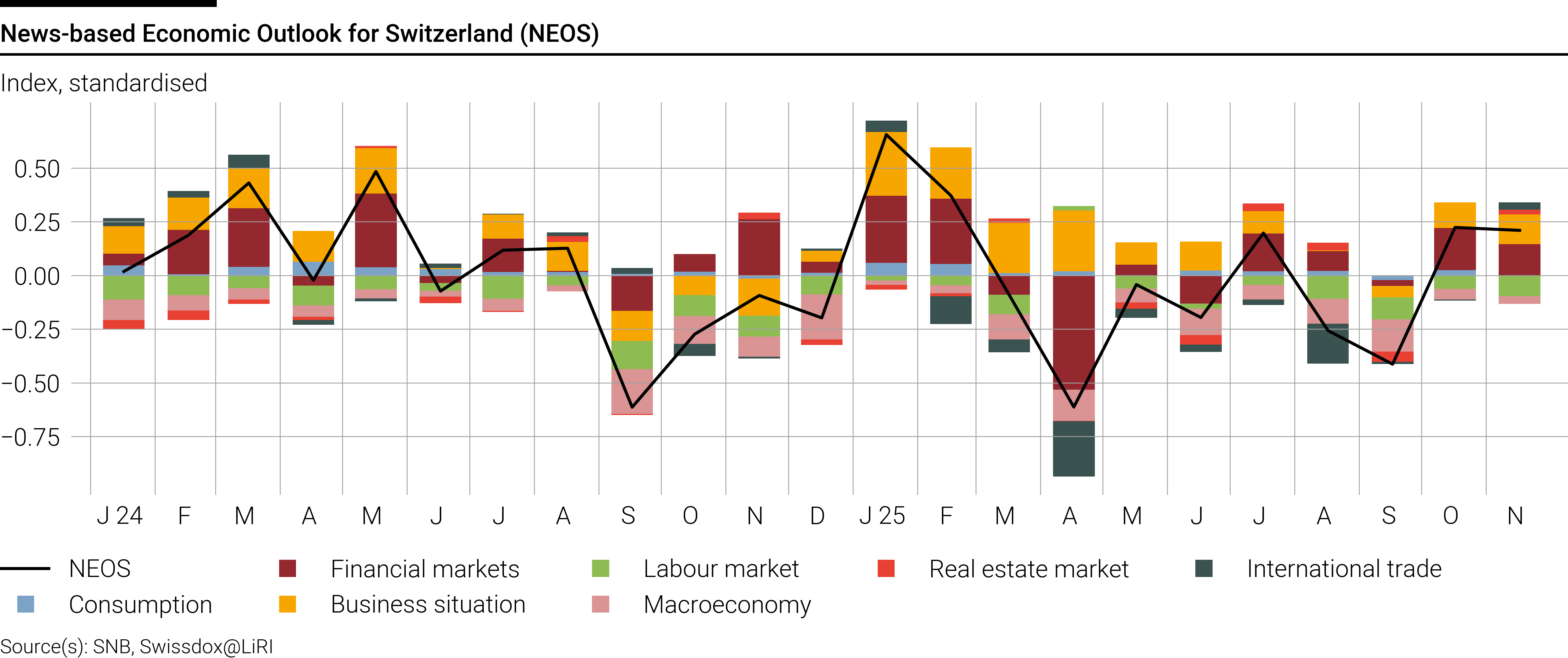}
    \caption{Contributions of pre-defined topics using the classification-based  approach.}
    \label{fig:contr_classifier}
\end{figure}

In Figure~\ref{fig:contr_classifier} we show the contributions of the topics determined using the classification-based  approach trained on the synthetic articles. In line with Figure~\ref{fig:neos_contr_keyword}, we observe a large negative contribution of trade-related articles in April and August~2025. Furthermore, in April~2025, the large negative contribution of financial market-related articles reflects the global stock market dip after ``Liberation Day''. Note that while some topics have only negative or positive contributions in the time range depicted in the figure, all topics contribute positively and negatively at some point in time over the whole time period.

\begin{figure}[ht]
    \centering
    \includegraphics[width=\textwidth]{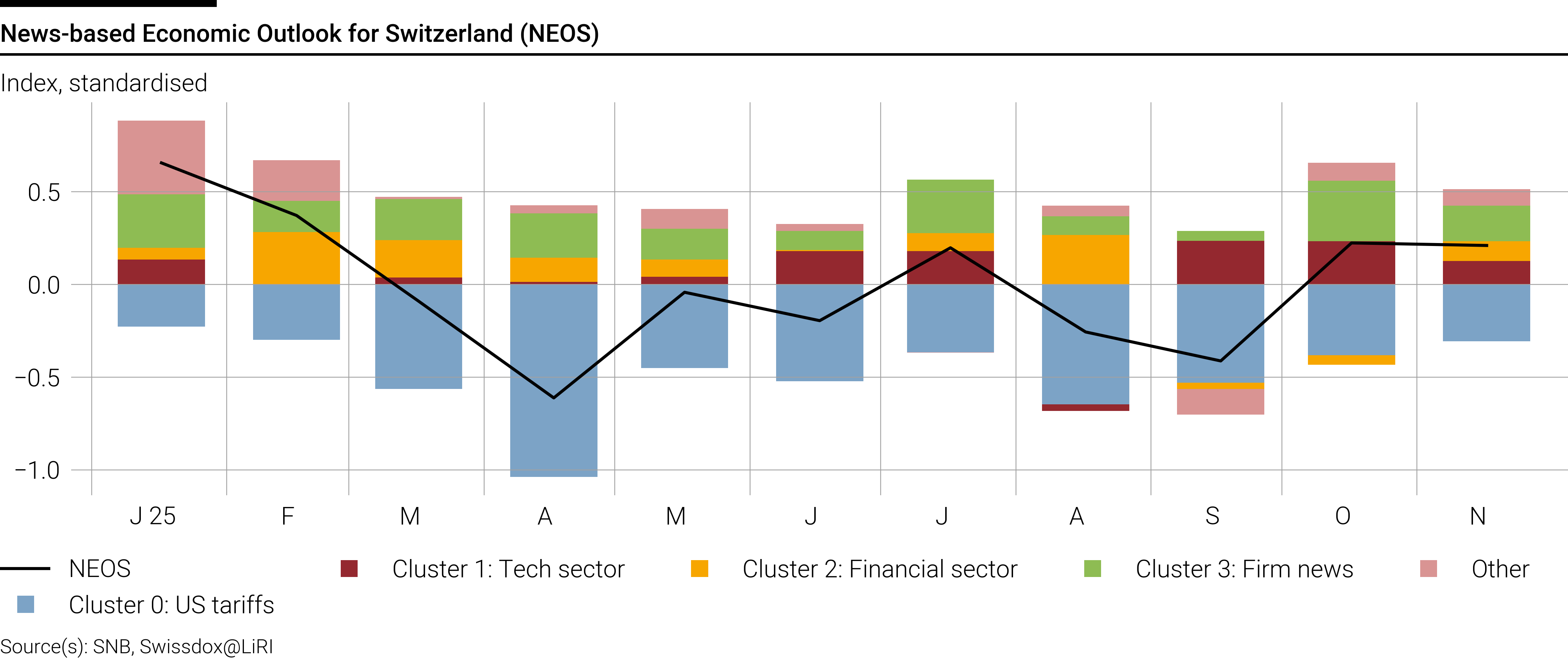}
    \caption{Contributions of topics identified using a clustering method.}
    \label{fig:contr_clustering}
\end{figure}

\begin{figure}[ht]
    \centering
    \begin{minipage}[b]{0.45\textwidth}
        \centering
        \includegraphics[width=\textwidth]{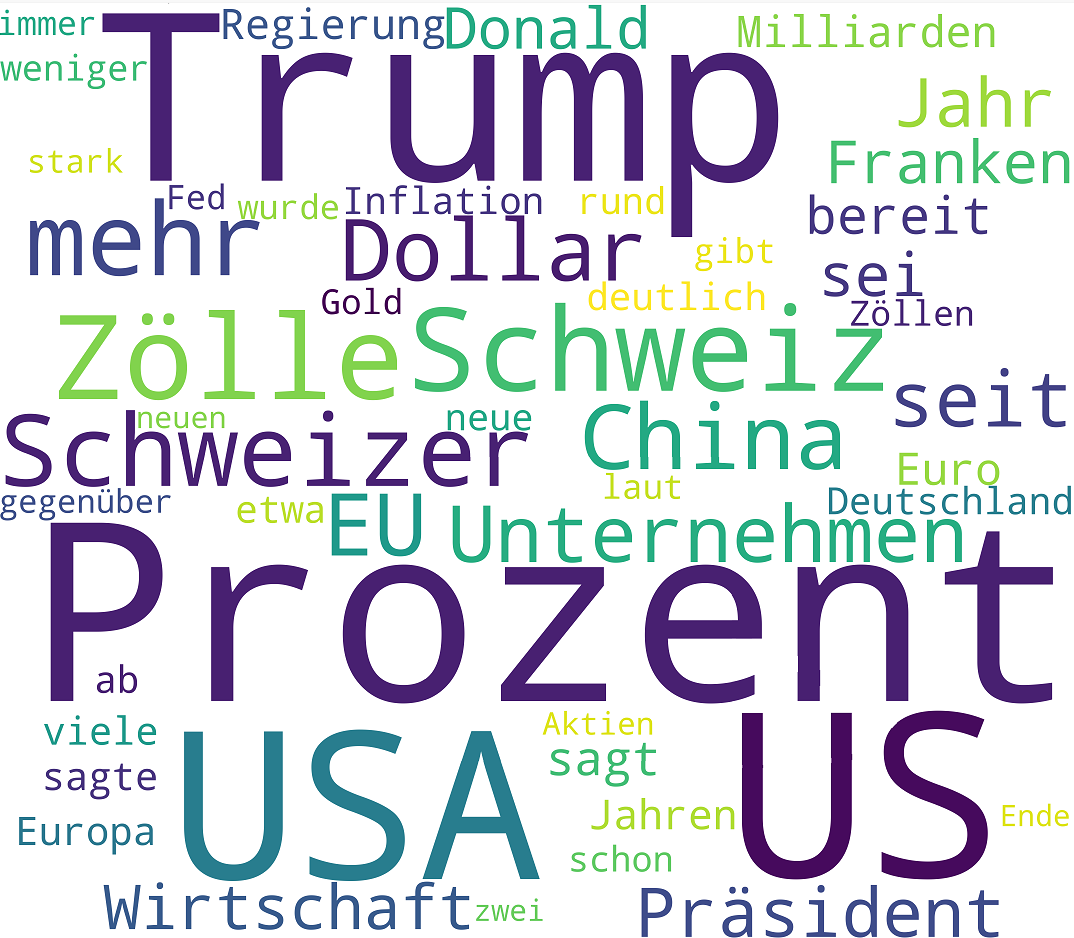}
        \caption{Word cloud for articles contained in Cluster 0.}
        \label{fig:wordcloud_cluster_0}
    \end{minipage}
    \hspace{0.5cm}
    \begin{minipage}[b]{0.45\textwidth}
        \centering
        \includegraphics[width=\textwidth]{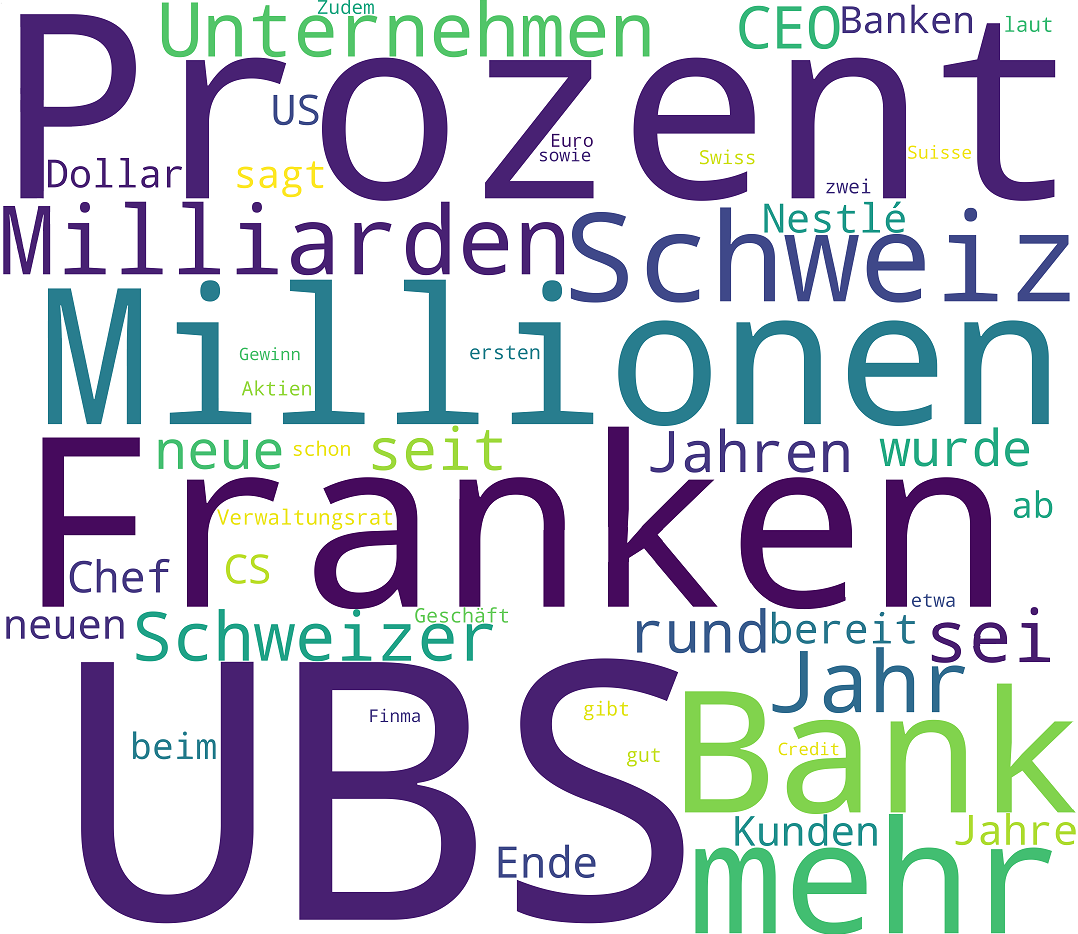}
        \caption{Word cloud for articles contained in Cluster 2.}
        \label{fig:wordcloud_cluster_2}
    \end{minipage}
    \label{fig:main_fig}
\end{figure}

In Figure~\ref{fig:contr_clustering} we depict the contributions of the topics identified using the clustering approach. We show the clusters for which the absolute value of the monthly contribution exceeds 0.2 at least once and summarize the other clusters in ``other''. In line with the previous two contribution results, the word cloud in Figure~\ref{fig:wordcloud_cluster_0} for Cluster~0 covers mainly topics related to the US's tariffs. Cluster~1 covers technology related developments, in the economy in general or of particular firms and contributes mostly positive. An LLM-generated summary of the headlines closest to the cluster center per month is the following: \textit{Overall, these titles depict Switzerland's dynamic role in the global tech ecosystem, emphasizing innovation, venture capital growth, international business expansion, and significant corporate moves (such as mergers and acquisitions). The Swiss tech sector seems to be benefiting from global trends, including the increasing prominence of start-ups, international investments, and the opportunities created by market instability elsewhere, especially in the US.} The word cloud for Cluster~2 in Figure~\ref{fig:wordcloud_cluster_2} shows mainly banking-related topics. The most negative article from October~2025 highlights how fears of a banking crisis in the US also weighed on Swiss financial stocks.\footnote{Our DUA with Swissdox@LiRI does not allow the original news articles to be published here as an illustration.} Cluster~3 includes articles about various companies and their business situation, i.e., headlines such as \textit{Quarterly results: Logitech exceeds expectations again in the second quarter} (translated from German to English).

\section{Sensitivity analysis}
\label{sec:sensitivity}

In the following, we present several sensitivity analyses to validate the key design choices of our methodology.

\subsection{Evaluation of the sentiment model}

To assess the quality of the predicted probabilities of the NEOS procedure (see Section~\ref{subsec:logreg}), we conducted an experiment with two human labelers who were assigned the same 150 news articles each. We randomly sample 50 articles each from three parts of the overall indicator distribution: 50 articles from the 1st-5th quantile (the lowest scores representing a negative outlook), 50 articles from the 45th-55th quantile (scores near the median representing a neutral outlook), and 50 articles from the 95th-100th quantile (the highest scores representing a positive outlook). The labelers were instructed to classify the articles into three categories: negative, neutral, and positive economic outlook.

The average alignment between the human and NEOS procedure labeling was 81\%, while the alignment between the two human labelers was 79\%. This implies that the labels generated using the NEOS procedure are at least as reliable as those produced by human labelers, and potentially more consistent. Neutral articles show the greatest divergence. Articles classified as neutral by our method were more frequently assigned to the positive or negative categories by human labelers. In contrast, the alignment for positive and negative articles was 89\%.

\subsection{Sensitivity to the number of synthetic articles}
\label{rem:neos_sample_size}

To assess whether 128 positive/128 negative synthetic articles are sufficient to ensure a stable sentiment score, we conduct an experiment where random subsamples of the synthetic articles are iteratively increased in size, and NEOS is computed for each subset. We found that using just 32 synthetic positive/negative articles resulted in high variance between the outcomes of different subsamples. However, as the number of synthetic articles increased, the variance decreased, with no further change after approximately 100 positive/negative synthetic articles. NEOS trained on subsets of approximately 100 synthetic positive/negative articles closely resembled the original NEOS results with 128 positive/128 negative synthetic articles. Increasing the number of synthetic articles further did not have any noticeable effect. Based on these findings, we conclude that a sample size of 128 positive/negative synthetic articles is sufficient for stable performance.

\subsection{Leakage of future information in the synthetic articles}
\begin{figure}[ht]
    \centering
    \includegraphics[width=\textwidth]{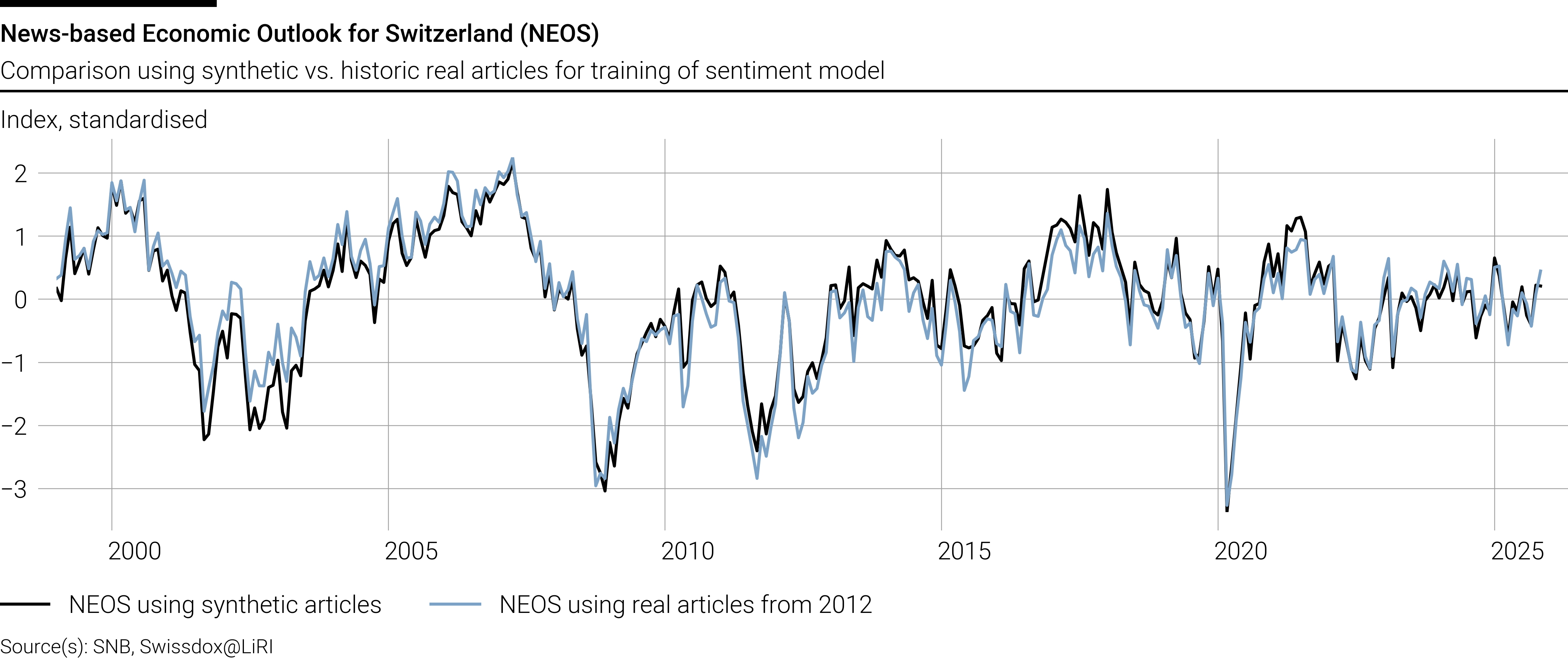}
    \caption{High correlation between the indicator obtained using news articles from 2012 and the one obtained using the synthetic articles.}
    \label{fig:neos_2012}
\end{figure}

Training data up to April 2024 was used for Claude 3.5 Sonnet, i.e., the LLM we use to generate the synthetic articles was trained on texts discussing past crises, such as the EU debt crisis or the COVID-19 pandemic. It is not a priori clear whether our indicator would capture the sentiment of future crises similarly well compared to past crises that were known during training. To show that NEOS does not depend on such future information, we construct the indicator by training the sentiment model on the embeddings of 256 real news articles from 2012 instead of the synthetic articles. We choose the articles by selecting the ones with the highest and lowest probability score from the originally trained logistic regression. The indicator obtained using real news articles from 2012 is shown together with the original monthly NEOS in Figure~\ref{fig:neos_2012}. The two indicators evolve in a similar way and have a correlation of 0.962 suggesting that our methodology does not rely on knowledge of future events in the synthetic articles.\footnote{This type of leakage is not unique to our indicator. Compound survey indicators (such as SECO-SEC and Swiss Economic Confidence) are based on multiple surveys. The composition of these indicators changes over time and their constituent subindices are selected based on historical correlations.}

\subsection{Relation of the logistic regression to cosine similarity}
\label{rem:cosine_logistic_connection}

During development, we explored an alternative sentiment scoring approach based on cosine similarity, which is often used for retrieval augmented generation (RAG). We generated the sentiment indicator by computing the cosine similarity between each relevant news article and each synthetic article and then averaging the similarity scores. This alternative approach measures how similar the relevant articles published in a given month are to the positive/negative synthetic articles. 

While this approach yields almost identical results, it is computationally much more costly because it requires computing $2\times N_{s}\times N$ similarity scores, where $N_{s}$ is the number of positive/negative synthetic articles and $N$ the total number of real, relevant news articles.

It is helpful to understand why the simpler regularized logistic regression works almost identically in this setting despite being far less resource-intense. 

The synthetic positive and negative articles form two well-separated clusters in the
embedding space (see Figure~\ref{fig:umap}). Let $p = \frac{1}{N_{s}}\sum_{i=1}^{N_{s}} p_i$ and $n = \frac{1}{N_{s}}\sum_{i=1}^{N_{s}} n_i$ denote the average embeddings of the synthetic positive and negative examples, respectively. The cosine-similarity–based score
\begin{equation*}
 \frac{1}{N_{s}}\sum_{i=1}^{N_{s}} \cos(x, p_i) -
\frac{1}{N_{s}}\sum_{i=1}^{N_{s}} \cos(x, n_i)
=\frac{1}{N_{s}}\sum_{i=1}^{N_{s}} x^\top(p_i-n_i) 
= x^\top(p-n),   
\end{equation*}
shows that projecting $x$ onto the vector $p-n$ measures whether $x$ aligns
more with the positive or negative cluster.

On the other hand, the logistic regression predicts
\begin{equation*}
\Pr(y=1 \mid x) = \sigma(x^\top w),
\quad 
\sigma(z)=\frac{1}{1+e^{-z}},
\end{equation*}
where $w$ denotes the fitted weight vector. In our empirical results, the raw NEOS
scores (before normalizing) translate into predicted probabilities close to $1/2$. In that region,
the sigmoid function is almost linear, and therefore
\begin{equation*}
\sigma(x^\top w) \approx \text{const.} + x^\top w'.
\end{equation*}
Thus, up to scaling and shifting, the logistic regression reduces to a linear score
based on a direction $w'$ that approximately corresponds to the separating
vector $p-n$. In other words, the model uses a projection similar to a
cosine-similarity-based approach. This provides additional affirmation that a
simple linear method suffices in our setting.

\section{Conclusion}
\label{sec:conclusion}

NEOS provides a complementary perspective to traditional survey-based indicators and demonstrates significant predictive power for forecasting GDP growth, outperforming traditional and lexicon-based indicators at different forecasting horizons. 
In contrast to traditional indicators, NEOS can be computed and updated in near real time, offering a timely advantage to indicators with fixed release schedules. This feature is especially valuable in times of crises.  

Our indicator is transparent and interpretable, which is key for policymakers. It is based on a single data source and if required, each article’s influence on the indicator can be measured and the content can be retrieved for interpretation. We also presented approaches to allow for systematic interpretability over time and to allow the detection of newly emerging topics. 

Our proposed methodology offers a way of leveraging state-of-the-art LLMs even if data are proprietary, which is usually the case when working with full text news articles, and on-premises resources are limited. Moreover, the structure of our approach is modular and flexible. We can adjust the constituent embedding model or LLM to keep track of technical developments. The modularity of our approach also makes it straightforward to build new indicators to track developments in other variables or to use different data sources.

\newpage

\bibliographystyle{plainnat}    
\bibliography{neos}

@TechReport{audrino:2024,
type={Working Papers},
institution={Swiss National Bank},
author={Francesco Audrino and Jessica Gentner and Simon Stalder},
title={Quantifying uncertainty: a new era of measurement through large language models},
year={2024},
number={2024-12},
doi={None},
url={https://ideas.repec.org/p/snb/snbwpa/2024-12.html},
}

@misc{Anthropic2024,
  author       = {{Anthropic}},
  title        = {Claude 3.5 Sonnet [Large language model]},
  howpublished = {\url{https://www.anthropic.com/news/claude-3-5-sonnet}},
  year         = {2024},
  note         = {Accessed: 2025-02-10}
}

@incollection{ash:2025,
  author       = {Elliott Ash and Stephen Hansen and Claudia Marangon and Yabra Muvdi},
  title        = {Large Language Models in Economics},
  booktitle    = {The Palgrave Handbook of Economics and Language},
  editor       = {Victor Ginsburgh and Shlomo Weber},
  publisher    = {Palgrave Macmillan},
  year         = {2025},
  note         = {Forthcoming (in press)},
}

@article{ashwin:2024,
author = {Ashwin, Julian and Kalamara, Eleni and Saiz, Lorena},
title = {Nowcasting Euro area GDP with news sentiment: A tale of two crises},
journal = {Journal of Applied Econometrics},
volume = {39},
number = {5},
pages = {887-905},
doi = {https://doi.org/10.1002/jae.3057},
url = {https://onlinelibrary.wiley.com/doi/abs/10.1002/jae.3057},
eprint = {https://onlinelibrary.wiley.com/doi/pdf/10.1002/jae.3057},
year = {2024}
}

@article{baker:bloom:davies:2016,
    author = {Baker, Scott R. and Bloom, Nicholas and Davis, Steven J.},
    title = {Measuring Economic Policy Uncertainty*},
    journal = {The Quarterly Journal of Economics},
    volume = {131},
    number = {4},
    pages = {1593-1636},
    year = {2016},
    month = {07},
    issn = {0033-5533},
    doi = {10.1093/qje/qjw024},
    url = {https://doi.org/10.1093/qje/qjw024},
    eprint = {https://academic.oup.com/qje/article-pdf/131/4/1593/30636768/qjw024.pdf},
}

@article{barbaglia:2025,
journal={Economic Inquiry},
author={Luca Barbaglia and Sergio Consoli and Sebastiano Manzan and Luca Tiozzo Pezzoli and Elisa Tosetti},
title={Sentiment analysis of economic text: A lexicon-based approach},
year={2025},
month={January},
pages={125-143},
volume={63},
number={1},
doi={10.1111/ecin.13264},
url={https://ideas.repec.org/a/bla/ecinqu/v63y2025i1p125-143.html},
}

@Article{beck:2026,
journal={Empirical Economics},
author={Elliot Beck and Michael Wolf},
title={Forecasting inflation with the hedged random forest},
year={2026},
month={February},
pages={1-36},
volume={70},
number={2},
doi={10.1007/s00181-025-02879-x},
url={https://ideas.repec.org/a/spr/empeco/v70y2026i2d10.1007_s00181-025-02879-x.html},
}

@techreport{binsbergen:2024,
 title = "(Almost) 200 Years of News-Based Economic Sentiment",
 author = "van Binsbergen, Jules H and Bryzgalova, Svetlana and Mukhopadhyay, Mayukh and Sharma, Varun",
 institution = "National Bureau of Economic Research",
 type = "Working Paper",
 series = "Working Paper Series",
 number = "32026",
 year = "2024",
 month = "January",
 doi = {10.3386/w32026},
 URL = "http://www.nber.org/papers/w32026",
}

@TechReport{buckmann:2025,
type={Bank of England working papers},
institution={Bank of England},
author={Marcus Buckmann and Ed Hill},
title={Improving text classification: logistic regression makes small {LLMs} strong and explainable tens-of-shot classifiers},
year={2025},
month={May},
number={1127},
doi={None},
url={https://ideas.repec.org/p/boe/boeewp/1127.html},
}

@article{garcia:2013,
author={Garc{\'\i}a, Diego},
title = {Sentiment during Recessions},
journal = {The Journal of Finance},
volume = {68},
number = {3},
pages = {1267-1300},
doi = {https://doi.org/10.1111/jofi.12027},
url = {https://onlinelibrary.wiley.com/doi/abs/10.1111/jofi.12027},
eprint = {https://onlinelibrary.wiley.com/doi/pdf/10.1111/jofi.12027},
year = {2013}
}

@TechReport{kwon:2025,
type={BIS Working Papers},
institution={Bank for International Settlements},
author={Byeungchun Kwon and Taejin Park and Phurichai Rungcharoenkitkul and Frank Smets},
title={Parsing the pulse: decomposing macroeconomic sentiment with {LLMs}},
year={2025},
month={Oct},
number={1294},
doi={None},
url={https://ideas.repec.org/p/bis/biswps/1294.html},
}

@article{kwon:2024,
title={Large Language Models: A Primer for Economists},
author={Byeungchun Kwon and Taejin Park and Fernando Perez‑Cruz and Phurichai Rungcharoenkitkul},
journal={BIS Quarterly Review},
year={2024},
month={December},
institution={Bank for International Settlements},
}

@article{bybee:2024,
author = {Bybee, Leland and Kelly, Bryan and Manela, Asaf and Xiu, Dacheng},
title = {Business News and Business Cycles},
journal = {The Journal of Finance},
volume = {79},
number = {5},
pages = {3105-3147},
doi = {https://doi.org/10.1111/jofi.13377},
url = {https://onlinelibrary.wiley.com/doi/abs/10.1111/jofi.13377},
eprint = {https://onlinelibrary.wiley.com/doi/pdf/10.1111/jofi.13377},
year = {2024}
}

@article{cessie:1992,
 ISSN = {00359254, 14679876},
 URL = {http://www.jstor.org/stable/2347628},
 author = {S. Le Cessie and J. C. Van Houwelingen},
 journal = {Journal of the Royal Statistical Society. Series C (Applied Statistics)},
 number = {1},
 pages = {191--201},
 publisher = {[Royal Statistical Society, Oxford University Press]},
 title = {Ridge Estimators in Logistic Regression},
 urldate = {2025-11-10},
 volume = {41},
 year = {1992}
}

@article{diebold:1995,
author = {Diebold, Francis and Mariano, Roberto},
year = {1995},
month = {07},
pages = {253-263},
title = {Comparing Predictive Accuracy},
volume = {13},
journal = {Journal of Business \& Economic Statistics},
doi = {10.1080/07350015.1995.10524599}
}

@misc{gao:2022,
      title={Precise Zero-Shot Dense Retrieval without Relevance Labels}, 
      author={Luyu Gao and Xueguang Ma and Jimmy Lin and Jamie Callan},
      year={2022},
      eprint={2212.10496},
      archivePrefix={arXiv},
      primaryClass={cs.IR},
      url={https://arxiv.org/abs/2212.10496}, 
}

@article{hirschenbuehl:2021,
journal={Economic Bulletin Articles},
author={Hirschb\"uhl, Dominik and Onorante, Luca and Saiz, Lorena},
title={Using machine learning and big data to analyse the business cycle},
year={2021},
month={None},
pages={None},
volume={5},
number={None},
doi={None},
url={https://ideas.repec.org/a/ecb/ecbart/202100052.html},
}

@article{kalamara:2022,
author = {Kalamara, Eleni and Turrell, Arthur and Redl, Chris and Kapetanios, George and Kapadia, Sujit},
title = {Making text count: Economic forecasting using newspaper text},
journal = {Journal of Applied Econometrics},
volume = {37},
number = {5},
pages = {896-919},
doi = {https://doi.org/10.1002/jae.2907},
url = {https://onlinelibrary.wiley.com/doi/abs/10.1002/jae.2907},
eprint = {https://onlinelibrary.wiley.com/doi/pdf/10.1002/jae.2907},
year = {2022}
}

@article{kirtac:2024,
title = {Sentiment trading with large language models},
journal = {Finance Research Letters},
volume = {62},
pages = {105227},
year = {2024},
issn = {1544-6123},
doi = {https://doi.org/10.1016/j.frl.2024.105227},
url = {https://www.sciencedirect.com/science/article/pii/S1544612324002575},
author = {Kemal Kirtac and Guido Germano}
}

@article{klahn:2025, 
title={From dictionaries to LLMs – an evaluation of sentiment analysis techniques for German language data}, 
volume={1}, 
DOI={10.1017/chr.2025.10005}, 
journal={Computational Humanities Research}, 
author={Kl\"ahn, Jannis and Borst-Graetz, Janos and Burghardt, Manuel}, 
year={2025}, 
pages={e4}}

@Article{loughran:2011,
journal={Journal of Finance},
author={Tim Loughran and Bill Mcdonald},
title={When Is a Liability Not a Liability? Textual Analysis, Dictionaries, and 10-Ks},
year={2011},
month={February},
pages={35-65},
volume={66},
number={1},
doi={j.1540-6261.2010.01625.x},
url={https://ideas.repec.org/a/bla/jfinan/v66y2011i1p35-65.html},
}

@misc{mcinnes:2020,
      title={{UMAP}: Uniform Manifold Approximation and Projection for Dimension Reduction}, 
      author={Leland McInnes and John Healy and James Melville},
      year={2020},
      eprint={1802.03426},
      archivePrefix={arXiv},
      primaryClass={stat.ML},
      url={https://arxiv.org/abs/1802.03426}, 
}

@inproceedings{sturua:2025,
author = {Sturua, Saba and Mohr, Isabelle and Kalim Akram, Mohammad and G\"{u}nther, Michael and Wang, Bo and Krimmel, Markus and Wang, Feng and Mastrapas, Georgios and Koukounas, Andreas and Wang, Nan and Xiao, Han},
title = {Jina Embeddings V3: Multilingual Text Encoder with Low-Rank Adaptations},
year = {2025},
isbn = {978-3-031-88719-2},
publisher = {Springer-Verlag},
address = {Berlin, Heidelberg},
url = {https://doi.org/10.1007/978-3-031-88720-8_21},
doi = {10.1007/978-3-031-88720-8_21},
booktitle = {Advances in Information Retrieval: 47th European Conference on Information Retrieval, ECIR 2025, Lucca, Italy, April 6–10, 2025, Proceedings, Part V},
pages = {123–129},
numpages = {7},
location = {Lucca, Italy}
}

@Article{seiler:2025,
journal={Economics Letters},
author={Seiler, Pascal},
title={Measuring economic sentiment from open-ended survey comments using large language models},
year={2025},
month={},
pages={},
volume={256},
number={C},
doi={10.1016/j.econlet.2025.112622},
url={https://ideas.repec.org/a/eee/ecolet/v256y2025ics0165176525004598.html},
}

@article{tetlock:2007,
author = {Tetlock, Paul C.},
title = {Giving Content to Investor Sentiment: The Role of Media in the Stock Market},
journal = {The Journal of Finance},
volume = {62},
number = {3},
pages = {1139-1168},
doi = {https://doi.org/10.1111/j.1540-6261.2007.01232.x},
url = {https://onlinelibrary.wiley.com/doi/abs/10.1111/j.1540-6261.2007.01232.x},
eprint = {https://onlinelibrary.wiley.com/doi/pdf/10.1111/j.1540-6261.2007.01232.x},
year = {2007}
}

@article{woloszko:2024,
title = {Nowcasting with panels and alternative data: The OECD weekly tracker},
journal = {International Journal of Forecasting},
volume = {40},
number = {4},
pages = {1302-1335},
year = {2024},
issn = {0169-2070},
doi = {https://doi.org/10.1016/j.ijforecast.2023.11.005},
url = {https://www.sciencedirect.com/science/article/pii/S0169207023001139},
author = {Nicolas Woloszko},
}

@misc{bucur:2025,
      title={All shocks are different: insights from sentiment and topic analysis using {LLMs}}, 
      author={Iulia Bucur and Ed Hill},
      year={2025},
      url={https://bankunderground.co.uk/2025/11/13/all-shocks-are-different-insights-from-sentiment-and-topic-analysis-using-llms/},
      pages={retrieved 2025-12-10}
}

@Article{schorfheide:song:2024,
journal={International Journal of Central Banking},
author={Frank Schorfheide and Dongho Song},
title={Real-Time Forecasting with a (Standard) Mixed-Frequency VAR During a Pandemic},
year={2024},
month={October},
pages={275-320},
volume={20},
number={4},
doi={None},
url={https://ideas.repec.org/a/ijc/ijcjou/y2024q4a5.html},
}

\newpage

\appendix

\renewcommand{\thesubsection}{\Alph{section}.\arabic{subsection}}
\renewcommand{\thetable}{\Alph{section}\arabic{table}}

\section*{Appendix}
\setcounter{section}{1}
\setcounter{subsection}{0}

\subsection{Example of a prompt and output}
\label{sec:prompt_examples}

As an example, to generate a synthetic news article covering financial markets, we used the prompt below in Claude 3.5 Sonnet. All prompts were formulated in such a way that the generated articles convey a positive or negative outlook/assessment.

\noindent
[Translated from German]

\begin{quote}
\small
I want to classify newspaper articles from the business sections of Swiss news (e.g., NZZ, Tagesanzeiger, Handelszeitung) according to whether they give a rather negative or positive outlook for developments on the financial markets. To do this, I use high-dimensional embeddings of the newspaper articles calculated with BERT. I then compare these embeddings with the embeddings of artificially generated newspaper articles that are prototypically positive or negative in terms of the outlook for the financial markets, using cosine similarity.

For this assignment, write three business articles each on financial and stock markets with a positive and negative outlook (six articles in total) that differ in content. Each article should be approximately 400-500 words long. These articles should simulate realistic business journalism and be diverse in content.

\textbf{Requirements:}
\begin{itemize}
    \item \textbf{Variety of topics:} Choose different topics related to the financial markets (e.g., stock and stock market developments (e.g., SMI, SPI, Dow Jones, DAX, etc.), bonds, commodities (e.g., gold, oil), and regulation in the financial sector).
    \item \textbf{Style \& structure:} The articles should be written like realistic journalistic articles, similar to texts from NZZ or Handelszeitung. Use a factual, fact-based writing style with typical business journalism phrasing.
    \item \textbf{Clear polarization:} The positive articles should present optimistic economic developments. The negative articles should highlight pessimistic developments. The economic perspective (positive or negative) must be clearly recognizable without appearing exaggerated or unrealistic.
    \item \textbf{Embedding-friendly wording:} Use key economic terms and technical terminology to achieve the highest possible selectivity of embeddings between positive and negative articles. Use a mixture of quantitative data (figures, statistics) and qualitative economic assessments. Vary sentence structures and wording between positive and negative articles to avoid bias in the classification.
\end{itemize}

These articles should serve as reference points to better classify real newspaper articles using cosine similarity.

\end{quote}

\noindent
\textbf{Sample text for positive outlook generated by Claude 3.5 Sonnet:}

\noindent
[Translated from German]

\begin{quote}
\small
Swiss stock market on track for record high: SMI breaks through important mark

The Swiss Market Index (SMI) continues its impressive upward trend, yesterday exceeding the psychologically important mark of 11,800 points for the first time. This positive development is being driven in particular by heavyweights Nestlé, Novartis, and Roche, all of which presented solid quarterly figures.

Analysts at Credit Suisse are optimistic about the further development of the Swiss benchmark index. “The fundamental data of the SMI companies are extremely robust, and we see further upside potential,” explains Martha Bernhard, chief strategist at the major bank. The defensive orientation of the index is proving to be an advantage in the current market phase.

The positive momentum is supported by persistently low inflation in Switzerland. With an inflation rate of 1.2 percent, Switzerland is well below the European average. This gives the Swiss National Bank (SNB) more monetary policy leeway than other central banks.

Small and medium-sized companies are also benefiting from this environment. The Swiss Performance Index (SPI) also reached a new all-time high, underscoring the broad base of the upturn. 
\end{quote}

\noindent
\textbf{Sample text for negative outlook generated by Claude 3.5 Sonnet:}

\noindent
[Translated from German]

\begin{quote}
\small
Investors concerned: Tech stocks experience historic sell-off

Technology stocks on Wall Street are currently experiencing their longest losing streak since the dot-com bubble. The technology-heavy Nasdaq index has plummeted by an alarming 28 percent since the beginning of the year. Former high-flyers have been hit particularly hard: shares in Meta, Facebook's parent company, have lost more than 60 percent of their value within six months.

Analysts at Credit Suisse are warning of further price losses. “The combination of rising interest rates, high inflation, and declining consumer spending is hitting the technology sector hard,” explains Thomas Müller, chief strategist at the major bank. The aggressive tightening of monetary policy by the US Federal Reserve is weighing particularly heavily on high-growth companies whose profits are only expected in the future.

The Swiss benchmark index SMI is also suffering from the weakness in tech. Shares in semiconductor manufacturer AMS Osram have lost over 40 percent since the beginning of the year. Market observers expect a sustained correction that could also spread to other sectors.
\end{quote}

Instead of using an LLM for text generation, we first experimented with short statements written by ourselves conveying positive or negative economic outlook, embedded them and used these embeddings to compute the indicator. We found, however, that our indicator improves significantly if the generated texts resemble real news articles in terms of length, style, structure etc. These findings are in line with \citet{gao:2022}.

\subsection{Correlations with GDP growth}
\label{sec:correlations}
Tables~\ref{tab:gdp_correlations_1999},~\ref{tab:gdp_correlations_2009} and~\ref{tab:gdp_correlations_2014} benchmark NEOS and its timelier versions against other survey-based indicators for economic outlook, and other text-based approaches introduced in Appendix~\ref{sec:lexicon}. The tables show the average correlations of the indicators with real year-on-year (yoy) and quarter-on-quarter (qoq) GDP growth for different lags over the periods Q1 1999 until Q1 2025, Q1 2009 until Q1 2025 and Q1 2014 until Q1 2025. Since GDP is a quarterly time series, we transform all monthly indicators to quarterly time series by taking three-month averages over the respective quarter. For instance, a lag of three quarters means that we compare GDP growth at time $t$ with the respective indicator at time $t-3$. 

NEOS correlates with GDP growth in a similar way to the other indicators and outperforms for certain lags. The correlations with quarter-on-quarter GDP growth are much lower for all indicators.\footnote{If qoq correlations with GDP growth are positive over multiple lags, yoy correlations will be higher by construction. Yoy GDP growth contains not only contemporaneous information but also information up to 12 months ago.} NEOS shows the highest correlation for a lag of one quarter with respect to year-on-year GDP. The contemporaneous correlations (lag 0) can still be interpreted as a lead of the indicator since GDP is released with a delay of two months.\footnote{The flash estimate is delayed by less than two months ($t + 45$ days.)} Note that Tables~\ref{tab:gdp_correlations_1999},~\ref{tab:gdp_correlations_2009} and~\ref{tab:gdp_correlations_2014} show ex post comparisons and do not account for actual data availability of the other indicators in the past and GDP revisions. Not all indicators are released immediately at the end of the month and historically, some indicators are only available quarterly.

\newpage
\begin{table}[H]
\caption{Analysis from Q1 1999 until Q1 2025}
    \label{tab:gdp_correlations_1999}

%\small
    \centering
    \begin{tabular}{l ccccc cc}
        \toprule
        & \multicolumn{5}{c}{\textbf{lag of indicator (quarters)}} & \multicolumn{2}{c}{\textbf{avg.\ over lags}} \\
        \cmidrule(lr){2-6} \cmidrule(lr){7-8}
        \textbf{Correlations with yoy GDP} & 0 & 1 & 2 & 3 & 4 & $0,1$ & $0, \ldots, 4$ \\
        \midrule
        NEOS & 0.60 & 0.68 & 0.64 & 0.51 & 0.24 & 0.64 & \textbf{0.53} \\
        NEOS (first 7 days in the month) & 0.66 & 0.65 & 0.59 & 0.42 & 0.14 & 0.65 & 0.49 \\
        NEOS (first 14 days in the month) & 0.66 & 0.66 & 0.60 & 0.43 & 0.14 & 0.66 & 0.50 \\
        NEOS (first 21 days in the month) & 0.65 & 0.66 & 0.61 & 0.44 & 0.15 & 0.66 & 0.50 \\
        Lexicon-based approach & 0.57 & 0.57 & 0.48 & 0.37 & 0.15 & 0.57 & 0.43 \\
        Manufacturing PMI for Switzerland & 0.74 & 0.74 & 0.62 & 0.41 & 0.14 & \textbf{0.74} & 0.53 \\
        SECO Consumer Sentiment Index & 0.46 & 0.41 & 0.35 & 0.22 & -0.01 & 0.43 & 0.29 \\

        %\midrule
        \addlinespace
        \addlinespace
        \textbf{Correlations with qoq GDP} & & & & & & & \\
        \midrule
        NEOS & 0.46 & 0.27 & 0.11 & 0.13 & -0.07 & \textbf{0.36} & \textbf{0.18} \\
        NEOS (first 7 days in the month) & 0.45 & 0.11 & 0.17 & 0.06 & -0.08 & 0.28 & 0.14 \\
        NEOS (first 14 days in the month) & 0.45 & 0.12 & 0.17 & 0.06 & -0.08 & 0.29 & 0.14 \\
        NEOS (first 21 days in the month) & 0.46 & 0.14 & 0.17 & 0.07 & -0.08 & 0.30 & 0.15 \\
        Lexicon-based approach & 0.37 & 0.19 & 0.01 & 0.10 & -0.03 & 0.28 & 0.13 \\
        Manufacturing PMI for Switzerland & 0.43 & 0.20 & 0.10 & 0.03 & -0.08 & 0.32 & 0.14 \\
        SECO Consumer Sentiment Index & 0.32 & 0.02 & 0.06 & -0.00 & -0.10 & 0.17 & 0.06 \\

        \bottomrule
    \end{tabular}
    
    \medskip
\raggedright
\footnotesize
Correlations with GDP growth for different lags. The highest average correlations are written in bold. \\
Data sources: KOF Swiss Economic Institute, procure.ch, SECO, SNB, Swissdox@LiRI, UBS.
\end{table}

\newpage
\begin{table}[H]
\caption{Analysis from Q1 2009 until Q1 2025}
    \label{tab:gdp_correlations_2009}

%\small
    \centering
    \begin{tabular}{l ccccc cc}
        \toprule
        & \multicolumn{5}{c}{\textbf{lag of indicator (quarters)}} & \multicolumn{2}{c}{\textbf{avg.\ over lags}} \\
        \cmidrule(lr){2-6} \cmidrule(lr){7-8}
        \textbf{Correlations with yoy GDP} & 0 & 1 & 2 & 3 & 4 & $0,1$ & $0, \ldots, 4$ \\
        \midrule
        NEOS & 0.48 & 0.50 & 0.42 & 0.30 & -0.03 & 0.49 & 0.33 \\
        NEOS (first 7 days in the month) & 0.56 & 0.47 & 0.38 & 0.18 & -0.15 & 0.52 & 0.29 \\
        NEOS (first 14 days in the month) & 0.55 & 0.48 & 0.40 & 0.20 & -0.14 & 0.52 & 0.30 \\
        NEOS (first 21 days in the month) & 0.55 & 0.49 & 0.40 & 0.22 & -0.13 & 0.52 & 0.30 \\
        Lexicon-based approach & 0.42 & 0.37 & 0.22 & 0.12 & -0.12 & 0.40 & 0.20 \\
        KOF Business Situation Indicator & 0.63 & 0.22 & -0.05 & -0.37 & -0.63 & 0.42 & -0.04 \\
        Manufacturing PMI for Switzerland & 0.68 & 0.66 & 0.52 & 0.33 & 0.05 & \textbf{0.67} & \textbf{0.45} \\
        SECO Consumer Sentiment Index & 0.37 & 0.29 & 0.23 & 0.11 & -0.11 & 0.33 & 0.18 \\

        %\midrule
        \addlinespace
        \addlinespace
        \textbf{Correlations with qoq GDP} & & & & & & & \\
        \midrule
        NEOS & 0.36 & 0.08 & -0.05 & 0.09 & -0.19 & \textbf{0.22} & 0.06 \\
        NEOS (first 7 days in the month) & 0.35 & -0.14 & 0.11 & -0.03 & -0.19 & 0.11 & 0.02 \\
        NEOS (first 14 days in the month) & 0.36 & -0.12 & 0.10 & -0.02 & -0.19 & 0.12 & 0.02 \\
        NEOS (first 21 days in the month) & 0.36 & -0.10 & 0.09 & -0.00 & -0.20 & 0.13 & 0.03 \\
        Lexicon-based approach & 0.23 & 0.04 & -0.14 & 0.04 & -0.13 & 0.14 & 0.01 \\
        KOF Business Situation Indicator & 0.14 & -0.38 & -0.17 & -0.21 & -0.26 & -0.12 & -0.18 \\
        Manufacturing PMI for Switzerland & 0.33 & 0.10 & 0.05 & 0.03 & -0.12 & 0.22 & \textbf{0.08} \\
        SECO Consumer Sentiment Index & 0.27 & -0.12 & 0.02 & -0.01 & -0.08 & 0.07 & 0.02 \\

        \bottomrule
    \end{tabular}
    
    \medskip
\raggedright
\footnotesize
Correlations with GDP growth for different lags. The highest average correlations are written in bold. \\
Data sources: KOF Swiss Economic Institute, procure.ch, SECO, SNB, Swissdox@LiRI, UBS.
\end{table}

\newpage
\begin{table}[H]
\caption{Analysis from Q1 2014 until Q1 2025}
    \label{tab:gdp_correlations_2014}
    \centering
    \begin{tabular}{l ccccc cc}
        \toprule
        & \multicolumn{5}{c}{\textbf{lag of indicator (quarters)}} & \multicolumn{2}{c}{\textbf{avg.\ over lags}} \\
        \cmidrule(lr){2-6} \cmidrule(lr){7-8}
        \textbf{Correlations with yoy GDP} & 0 & 1 & 2 & 3 & 4 & $0,1$ & $0, \ldots, 4$ \\
        \midrule
        NEOS & 0.48 & 0.52 & 0.47 & 0.38 & -0.03 & 0.50 & 0.37 \\
        NEOS (first 7 days in the month) & 0.56 & 0.46 & 0.44 & 0.26 & -0.18 & 0.51 & 0.31 \\
        NEOS (first 14 days in the month) & 0.55 & 0.47 & 0.45 & 0.27 & -0.17 & 0.51 & 0.32 \\
        NEOS (first 21 days in the month) & 0.55 & 0.48 & 0.46 & 0.29 & -0.15 & 0.51 & 0.33 \\
        Lexicon-based approach & 0.43 & 0.40 & 0.26 & 0.17 & -0.13 & 0.41 & 0.23 \\
        KOF Business Situation Indicator & 0.63 & 0.19 & -0.03 & -0.35 & -0.65 & 0.41 & -0.04 \\
        Manufacturing PMI for Switzerland & 0.67 & 0.64 & 0.54 & 0.40 & 0.11 & \textbf{0.66} & \textbf{0.47} \\
        Service PMI for Switzerland & 0.67 & 0.48 & 0.29 & 0.16 & -0.32 & 0.58 & 0.26 \\
        SECO Consumer Sentiment Index & 0.39 & 0.26 & 0.22 & 0.13 & -0.11 & 0.33 & 0.18 \\
        
        %\midrule
        \addlinespace
        \addlinespace
        \textbf{Correlations with qoq GDP} & & & & & & & \\
        \midrule
        NEOS & 0.43 & 0.09 & -0.06 & 0.14 & -0.23 & \textbf{0.26} & 0.07 \\
        NEOS (first 7 days in the month) & 0.44 & -0.19 & 0.16 & -0.01 & -0.25 & 0.12 & 0.03 \\
        NEOS (first 14 days in the month) & 0.44 & -0.17 & 0.14 & -0.06 & -0.26 & 0.14 & 0.03 \\
        NEOS (first 21 days in the month) & 0.44 & -0.14 & 0.13 & -0.07 & -0.26 & 0.15 & 0.04 \\
        Lexicon-based approach & 0.29 & 0.06 & -0.18 & 0.07 & -0.15 & 0.18 & 0.02 \\
        KOF Business Situation Indicator & 0.18 & -0.41 & -0.15 & -0.20 & -0.27 & -0.12 & -0.17 \\
        Manufacturing PMI for Switzerland & 0.36 & 0.12 & 0.07 & 0.07 & -0.11 & 0.24 & \textbf{0.10} \\
        Service PMI for Switzerland & 0.50 & -0.09 & -0.15 & -0.04 & -0.25 & 0.20 & -0.01 \\
        SECO Consumer Sentiment Index & 0.29 & -0.14 & 0.04 & 0.00 & -0.07 & 0.08 & 0.02 \\

        \bottomrule
    \end{tabular}
    
    \medskip
\raggedright
\footnotesize
Correlations with GDP growth for different lags. The highest average correlations are written in bold. \\
Data sources: KOF Swiss Economic Institute, procure.ch, SECO, SNB, Swissdox@LiRI, UBS.
\end{table}

\subsection{EPU and lexicon-based approach to measure sentiment from news articles}
\label{sec:lexicon}
We compute two alternative text-based indicators, which are based on the same underlying news articles data as NEOS. 

a) We closely follow \citet{baker:bloom:davies:2016} to construct an EPU index for Switzerland, using their keywords plus Switzerland all translated to German and French.

b) To construct an alternative text-based sentiment indicator regarding the economic outlook we apply a lexicon-based approach where one relies on a predefined list of words that are associated with positive, negative, or neutral sentiment (see, e.g., \citet{loughran:2011}). Occurrences of these words are counted in a text and a sentiment score based on these occurrences is assigned. We use the lexicon developed by \cite{barbaglia:2025}. The lexicon is in English and consists of more than six thousand words. We translated and adapted it to German and applied it to the German news articles.\footnote{All articles in German that enter the computation of NEOS are used.} The results of the lexicon-based approach are not very promising. Simple word-count techniques face challenges in recognizing context and detecting negations in texts, for example, “not happy” (see, e.g., \citet{binsbergen:2024}). We could also translate the lexicon to French, but we refrain from the additional effort as the German articles are a clear majority in our sample and in an additional computing exercise, we observed that NEOS with and without the French articles looks very similar. Thus, the French articles do not seem to contain much additional information for sentiment, and this should also hold applying other approaches such as the lexicon-based one.

\subsection{Forecast evaluation for subperiods}
\label{sec:forecast_eval_subperiods}

\begin{table}[H]
\centering
\caption{
Forecasting Swiss year-on-year GDP growth from Q1~2009 to Q1~2025}
 \label{tab:MAE_2009}
\begin{tabular}{lccc}
\toprule
\textbf{Indicator} & \textbf{$h=0$} & \textbf{$h=1$} & \textbf{$h=2$} \\
\midrule
NEOS & 0.94 (0.18) & 1.00 (0.51) & 0.93 (0.10) \\
NEOS (first 7 days in the third month) & \textbf{0.93} (0.17) & 0.99 (0.44) & 0.93 (0.09) \\
NEOS (first 14 days in the third month) & 0.94 (0.21) & 0.98 (0.43) & 0.94 (0.12) \\
NEOS (first 21 days in the third month) & 0.95 (0.25) & \textbf{0.98} (0.41) & 0.94 (0.13) \\
\midrule
Lexicon-based approach & 1.20 (0.95) & 1.11 (0.74) & 1.05 (0.68) \\
EPU for Switzerland & 1.12 (0.86) & 1.11 (0.89) & 1.01 (0.52) \\
Manufacturing PMI for Switzerland & 1.03 (0.61) & 1.01 (0.53) & \textbf{0.90} (0.20) \\
KOF Business Situation Indicator & 1.06 (0.86) & 1.00 (0.67) & 1.00 (0.48) \\
SECO Consumer Sentiment Index & 1.20 (0.79) & 1.07 (0.76) & 1.00 (0.52) \\
\bottomrule
\end{tabular}

\medskip
\raggedright
\footnotesize
MAE ratios with the Diebold–Mariano p-values in parentheses. The best MAE values are in bold. \\
Data sources: KOF Swiss Economic Institute, procure.ch, SECO, SNB, Swissdox@LiRI, UBS. \\
\medskip
\end{table}

\newpage
\begin{table}[H]
\centering
\caption{
Forecasting Swiss year-on-year GDP growth from Q1~2014 to Q1~2025}
 \label{tab:MAE_2014}
\begin{tabular}{lccc}
\toprule
\textbf{Indicator} & \textbf{$h=0$} & \textbf{$h=1$} & \textbf{$h=2$} \\
\midrule
NEOS & 0.94 (0.20) & 0.91 (0.09) & 0.89 (0.04) \\
NEOS (first 7 days in the third month) & 0.94 (0.22) & 0.91 (0.11) & 0.88 (0.04) \\
NEOS (first 14 days in the third month) & 0.95 (0.28) & 0.91 (0.13) & 0.88 (0.06) \\
NEOS (first 21 days in the third month) & 0.97 (0.33) & \textbf{0.90} (0.12) & \textbf{0.87} (0.06) \\
\midrule
Lexicon-based approach & 1.29 (0.97) & 1.06 (0.60) & 1.05 (0.62) \\
EPU for Switzerland & 0.99 (0.49) & 1.04 (0.88) & 0.99 (0.45) \\
Manufacturing PMI for Switzerland & 1.05 (0.64) & 0.84 (0.24) & 0.87 (0.29) \\
KOF Business Situation Indicator & 1.05 (0.63) & 1.01 (0.63) & 0.97 (0.30) \\
SECO Consumer Sentiment Index & 1.10 (0.62) & 1.00 (0.51) & 1.05 (0.70) \\
Service PMI for Switzerland\textsuperscript & \textbf{0.86} (0.12) & 0.94 (0.16) & 0.93 (0.09) \\
\bottomrule
\end{tabular}

\medskip
\raggedright
\footnotesize
MAE ratios with the Diebold–Mariano p-values in parentheses. The best MAE values are in bold. \\
Data sources: KOF Swiss Economic Institute, procure.ch, SECO, SNB, Swissdox@LiRI, UBS. \\
\medskip
\end{table}

\newpage
\subsection{Cumulative absolute error differences for all horizons}
\label{sec:caed}
\begin{figure}[ht!]
    \centering
     \includegraphics[width=\textwidth]{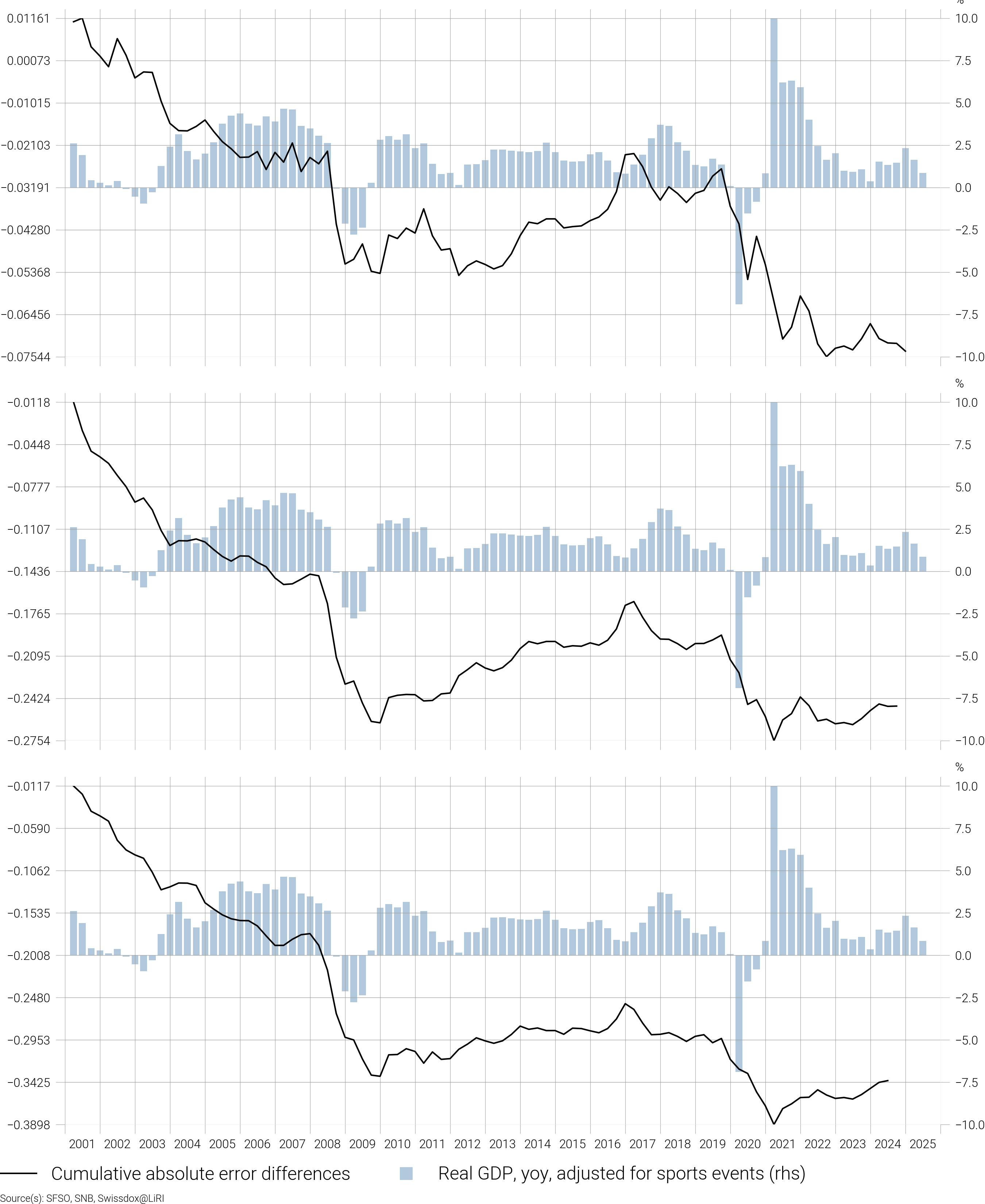}
    \caption{Cumulative forecast improvements compared to an AR(1) model for forecast horizons $h=0$ (top), $h=1$ (middle) and $h=2$ (bottom).}
    \label{fig:crisis_appendix}
 \end{figure}
 
\newpage
\subsection{Accuracy vs.\ precision for different candidate models to classify relevant articles}
\label{sec:accuracy vs precision}
\begin{figure}[ht!]
\centering
\includegraphics[width=\textwidth]{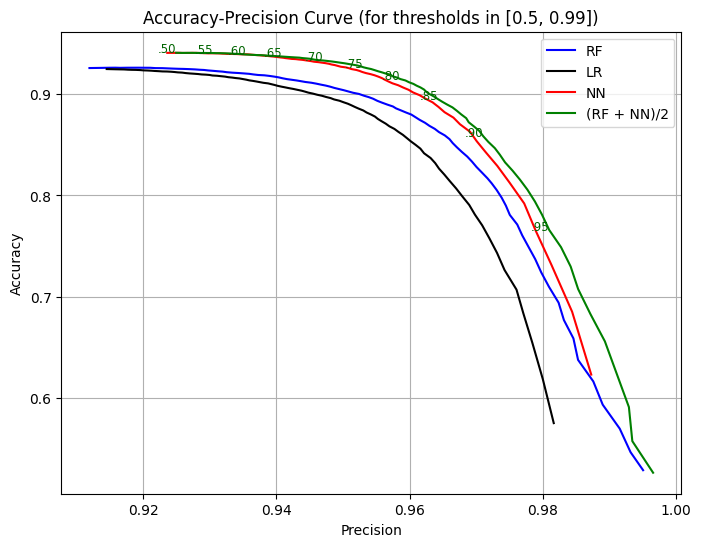}
\caption{Accuracy vs.\ precision for different candidate models to classify relevant articles. We use the abbreviations random forest (RF), L2-regularized logistic regression (LR), and ReLU feed-forward neural network (NN). (RF + NN)/2 is the equally-weighted ensemble. Every point in the plot corresponds to a particular threshold $\tau \in [0.5, 0.99]$. We assign label 1 whenever the predicted probability is larger than $\tau$.}
\label{fig:acc_prec}
\end{figure}

\end{spacing}
\end{document}